\documentclass[twocolumn,superscriptaddress,amsfont,amssymb,amsmath, showpacs,balancelastpage, nofootinbib, preprintnumbers]{revtex4-2}
\usepackage{graphicx,longtable,mathrsfs,color,array}
\usepackage[unicode=true,pdfusetitle,
bookmarks=true,bookmarksnumbered=true,bookmarksopen=true,bookmarksopenlevel=1,
breaklinks=false,pdfborder={0 0 0},backref=false,colorlinks=true]{hyperref}
\hypersetup{citecolor=blue,filecolor=blue,linkcolor=blue,urlcolor=blue}
\usepackage[usenames,dvipsnames]{xcolor}
\usepackage{amssymb,amsmath,mathtools,mathrsfs,enumitem}
\usepackage{epsfig,subfigure,placeins,float}
\usepackage{booktabs,longtable,multirow}
\usepackage{exscale,relsize}
\usepackage[normalem]{ulem}
\usepackage[T1]{fontenc}
\usepackage[utf8]{inputenc}
\usepackage{enumerate}
\usepackage{times, mathptmx}
\DeclareMathAlphabet{\mathcm}{OT1}{cmr}{m}{it}
\renewcommand{\v}{\mathcm{v}}

\usepackage{tikz}
\usepackage[compat=1.1.0]{tikz-feynman}
\usetikzlibrary{arrows,decorations.markings}
\usepackage{soul}

\usepackage{breqn}

\usetikzlibrary{decorations.pathmorphing}

\usetikzlibrary{arrows,positioning,decorations.markings,decorations.pathmorphing,calc}
\definecolor{hyperref}{RGB}{026,028,087}

\usepackage{tikz}
\usepackage[compat=1.1.0]{tikz-feynman}

\newcommand{\gravconv}{
\begin{tikzpicture}[baseline]
\begin{feynman}
\vertex (a);
\vertex [right=1cm of a] (b);
\diagram* {
(a) -- [photon] (b)
}; 
\end{feynman}
\end{tikzpicture}
}

\newcommand{\TidalHconv}{
\begin{tikzpicture}[baseline]
\begin{feynman}
\vertex [crossed dot, minimum size=0.25cm, line width=1.5pt](a) {};
\vertex [right=1cm of a] (b);
\diagram* {
(a) -- [plain, line width=1.5pt] (b)
}; 
\end{feynman}
\end{tikzpicture}
}

\newcommand{\MassSc}{
\begin{tikzpicture}[baseline]
\begin{feynman}
\vertex [dot, minimum size=0.25cm, label=180:$m$] (a) {};
\vertex [right=0.8cm of a] (c);
\vertex [right=0.2cm of c] (d);
\diagram* {
(a) -- [photon] (c),
(c) -- [plain, -stealth] (d)
}; 
\end{feynman}
\end{tikzpicture}
}

\newcommand{\LinT}{
\begin{tikzpicture}[baseline]
\begin{feynman}
\vertex [square dot, minimum size=0.25cm, label=180:$\lambda_2$] (a) {};
\vertex [below=0.8cm of a, crossed dot, minimum size=0.25cm, line width=1.5pt] (aT) {};
\vertex [right=0.6cm of a] (d);
\vertex [right=0.2cm of d] (d2);
\diagram* {
(a) -- [photon] (d),
(d) -- [plain, -stealth] (d2),
(aT) -- [plain, line width=1.5pt] (a)
}; 
\end{feynman}
\end{tikzpicture}
}

\newcommand{\PotConv}{
\begin{tikzpicture}[baseline]
\begin{feynman}
\vertex (a);
\vertex [right=1cm of a] (b);
\diagram* {
(a) -- [scalar] (b)
}; 
\end{feynman}
\end{tikzpicture}
}

\usepackage[stretch=20,shrink=10,final]{microtype}
\let\revtitle\maketitle
\renewcommand{\maketitle}{%
	\revtitle
	\tolerance=500
	\hyphenpenalty=1000
}

\newcommand{\horizontalrule}[1]{\noindent\hfil\rule{#1}{0.5pt}}

\makeatletter
\def\l@@dottedsections#1#2#3#4{%
	\begingroup
	\everypar{}%
	\set@tocdim@pagenum\@tempboxa{#4}%
	\global\@tempdima\csname tocdim@#2\endcsname
	\leftskip\csname tocleft@#2\endcsname\relax
	\dimen@\csname tocleft@#1\endcsname\relax
	\parindent-\leftskip\advance\parindent\dimen@
	\rightskip\tocleft@pagenum plus 1fil\relax
	\skip@\parfillskip\parfillskip\z@
	\let\numberline\numberline@@sections
	\@nameuse{l@f@#2}%
	\ignorespaces#3\unskip\nobreak
	\leaders \hbox {$\m@th \mkern \@dotsep mu\hbox {.}\mkern \@dotsep mu$}
	\hskip\skip@
	\hb@xt@\rightskip{\hfil\unhbox\@tempboxa}\hskip-\rightskip\hskip\z@skip
	\expandafter\par
	\expandafter\aftergroup\csname tocdim@#2%
	\expandafter\endcsname
	\expandafter\endgroup
	\the\@tempdima\relax
}
\def\l@subsection{\l@@dottedsections{section}{subsection}}
\def\l@subsubsection{\l@@dottedsections{subsection}{subsubsection}}

\makeatother

\makeatletter

\renewcommand*\l@section[2]{%
  \ifnum\c@tocdepth>0\relax
    \addpenalty\@secpenalty
    \addvspace{0.8em plus 0.2em minus 0.2em}%
    \@dottedtocline{1}{-1em}{1.5em}{#1}{#2}%
  \fi}

\makeatother

\let\revtoc\tableofcontents
\makeatletter
\renewcommand{\tableofcontents}{%
	\onecolumngrid%
	\vskip -7pt%
	\horizontalrule{\textwidth}%
	\vskip 10pt%
	\bgroup
	\twocolumngrid%
	\raggedbottom
	\revtoc
	\onecolumngrid%
	\egroup
	\vskip 10pt%
	\horizontalrule{\textwidth}%
	\vskip 20pt plus 5pt%
	\twocolumngrid%
}
\makeatletter

\usepackage{etoolbox}

\def\gsim{ \lower .75ex \hbox{$\sim$} \llap{\raise .27ex \hbox{$>$}} }
\def\lsim{ \lower .75ex \hbox{$\sim$} \llap{\raise .27ex \hbox{$<$}} }
\def\be{\begin{equation}}
\def\ee{\end{equation}}
\def\bea{\begin{eqnarray}}
\def\eea{\end{eqnarray}}

\def\Mm{{\cal M}}
\def\rorb{{r_{\rm o}}}

\DeclareMathOperator{\e}{e}

\newcommand{\comment}[1]{}
\newcommand{\Mpl}{M_{\rm Pl}}

\newcommand{\Hzerf}{{}^{(1)} \! H_0}
\newcommand{\Honef}{{}^{(1)} \! H_1}
\newcommand{\Htwof}{{}^{(1)} \! H_2}
\newcommand{\Hf}{{}^{(1)} \! H}
\newcommand{\Hs}{{}^{(2)} \! H}
\newcommand{\Kf}{{}^{(1)} \! K}
\newcommand{\Rf}{{}^{(1)} \! R}
\newcommand{\Vf}{{}^{(1)} \! V}
\newcommand{\deltaf}{{}^{(1)} \! \delta}
\newcommand{\Ef}{{}^{(1)} \! E}
\newcommand{\Hzers}{{}^{(2)} \! H_0}
\newcommand{\Hones}{{}^{(2)} \! H_1}

\newcommand{\Vs}{{}^{(2)} \! V}
\newcommand{\deltas}{{}^{(2)} \! \delta}
\newcommand{\Es}{{}^{(2)} \! E}
\newcommand{\Rs}{R_{*}}
\newcommand{\Rsec}{{}^{(2)} \! R}
\newcommand{\blin}{a}
\newcommand{\cupa}{c_\uparrow}
\newcommand{\cdown}{c_\downarrow}

\def\D{{\rm d}}
\def\dd{{\rm d}}

\definecolor{Gray}{gray}{0.9}
\definecolor{LightCyan}{rgb}{0.88,1,1}

\renewcommand*{\mathcolor}{}
\def\mathcolor#1#{\mathcoloraux{#1}}
\newcommand*{\mathcoloraux}[3]{%
\protect\leavevmode
\begingroup
color#1{#2}#3%
\endgroup
}

\newcommand{\textins}[2][fu-grey]{
\ifmmode\mathcolor{#1}{#2}
\else\textcolor{#1}{#2}\@\,
\fi
}
\graphicspath{{./}}
\allowdisplaybreaks

\tikzstyle{vecArrow} = [thick, decoration={markings,mark=at position
1 with {\arrow[semithick]{open triangle 60}}},
double distance=1.4pt, shorten >= 5.5pt,
preaction = {decorate},
postaction = {draw,line width=1.4pt, white,shorten >= 4.5pt}]

\makeatletter
\let\cat@comma@active\@empty
\makeatother

\begin{document}


\hypersetup{pageanchor=false} 
\title{Nonlinear Relativistic Tidal Response of Neutron Stars}

\author{Paolo Pani}
\affiliation{Dipartimento di Fisica, ``Sapienza'' Universit\`a di Roma \& Sezione INFN Roma1, P.A.~Moro 5, 00185, Roma, Italy}

\author{Massimiliano Maria Riva}
\affiliation{Department of Physics, Center for Theoretical Physics, Columbia University, New York, 538 West 120th Street, NY 10027, U.S.A.}
\author{Luca Santoni}
\affiliation{Universit\'e Paris Cit\'e, CNRS, Astroparticule et Cosmologie, 10 Rue Alice Domon et L\'eonie Duquet, F-75013 Paris, France}
\author{Nikola Savi\'c}
\affiliation{Universit\'e Paris-Saclay, CNRS, CEA, Institut de Physique Th\'eorique, 91191 Gif-sur-Yvette, France}
\author{Filippo Vernizzi}
\affiliation{Universit\'e Paris-Saclay, CNRS, CEA, Institut de Physique Th\'eorique, 91191 Gif-sur-Yvette, France}
\affiliation{Yukawa Institute for Theoretical Physics,
Kyoto University, 606-8502, Kyoto, Japan}

\begin{abstract}
\noindent 
We investigate the nonlinear tidal response of relativistic neutron stars by computing the fully relativistic, static, quadratic Love numbers. Using both the worldline effective field theory for extended gravitating bodies and second-order perturbations of relativistic stellar models, we derive the nonlinear tidal deformation induced by an external gravito-electric tidal field to quadratic order.
Through a suitable matching procedure, we provide for the first time the leading nonlinear tidal corrections to the conservative dynamics and gravitational-wave signal of binary systems. Quadratic Love numbers are
enhanced more than the linear ones in the small-compactness limit. Because of this, despite entering the gravitational-wave phase at 8th post-Newtonian (PN) order, the leading quadratic Love number can be as important as the next-to-next-to-leading order linear tidal correction, which enters at 7th PN order, and is larger than the subleading point-particle contribution entering at 4th PN order. In particular, quadratic Love numbers can be as large as $\sim 10\%$ of the linear Love numbers in the late inspiral phase.
Our approach provides a  framework to also compute the (subleading) nonlinear effects induced by magnetic tidal fields and higher multipole moments, and sets the foundations for incorporating nonlinear tidal effects in high-precision gravitational-wave modeling.
\end{abstract}


\preprint{YITP-26-32}

\date{\today}
\maketitle

\tableofcontents
\section{Introduction} 

Neutron stars are fascinating and complex astrophysical objects~\cite{Lattimer:2021emm}. Their interiors host rare phenomena and extreme conditions not observed elsewhere, with cores containing matter at the highest densities in the observable universe. This makes them unique  laboratories for probing the behavior of dense matter, providing connections across nuclear physics, particle physics, and astrophysics.

The interior structure of a neutron star is commonly described in terms of an equation of state (EoS), which relates the internal density, pressure, and temperature. This relation, which depends sensitively on the microscopic details of the star's constituents~\cite{Lattimer:2021emm,Burgio:2021vgk}, uniquely determines its macroscopic properties, including the mass-radius relationship, maximum mass, size, and tidal deformability. Through precise measurements of the gravitational-wave (GW) signal emitted during the coalescence of merging binary systems, GW astronomy offers a powerful window into these properties, providing  an unprecedented opportunity to explore subatomic physics under conditions previously inaccessible~\cite{Chatziioannou:2020pqz}.

When a neutron star is subjected to an external tidal gravitational field, it deforms, developing induced multipole moments~\cite{Poisson_Will_2014,Hinderer:2007mb,Flanagan:2007ix,Damour:2009vw,Binnington:2009bb}. This response, commonly parameterized by the so-called Love numbers, affects the system's binding energy and GW flux, leaving a measurable imprint on the emitted GWs and providing direct information about the star's internal structure. In this context, most studies have focused on the \emph{linear} regime of tidal effects, wherein the induced moments of one binary component are proportional to the tidal field produced by its companion~\cite{Hinderer:2007mb,Damour:2009vw,Binnington:2009bb,Landry:2015cva,Landry:2015zfa,Pani:2015hfa,Pani:2015nua,Landry:2017piv,Poisson:2020mdi}. Although linear tidal effects dominate the early inspiral phase of neutron star binary coalescences, it is both theoretically and observationally motivated to understand and quantify how \emph{nonlinearities} influence the gravitational waveform. First, nonlinear corrections become increasingly important in the late inspiral, where the dynamics of the system approaches the nonlinear regime and the adiabatic approximation breaks down.  In the light of future high-precision GW observations, incorporating these effects may therefore be essential for obtaining accurate waveform templates and avoiding systematic biases~\cite{Gamba:2020wgg,JimenezForteza:2018rwr,Gupta:2024gun,ET:2025xjr}. In addition, even though nonlinear Love numbers first enter the GW phase at 8th order of the post-Newtonian (PN) expansion~\cite{Blanchet:2013haa}---similar to their dynamical counterparts (see, e.g.,~\cite{Chakrabarti:2013lua,Hinderer:2016eia,Steinhoff:2016rfi,Poisson:2020vap,Mandal:2023hqa,Pitre:2023xsr,Katagiri:2024wbg,HegadeKR:2024agt,Chakraborty:2025wvs,Combaluzier--Szteinsznaider:2025eoc, Kobayashi:2025vgl})---they can receive an enhancement in powers of the ratio of the star's radius to the Schwarzschild radius for less compact stars. Furthermore, understanding the role of nonlinearities is  key to breaking observational degeneracies and uncovering the fundamental origin of emergent symmetries and universal behaviors of neutron stars~\cite{Yagi:2013bca,Yagi:2016bkt,Kunz:2022wnj}. 

In this work we ask a simple question: \textit{How do field nonlinearities affect the tidal deformability of a relativistic neutron star and the GW signals from neutron-star binaries?}
Nonlinear tidal responses have been previously studied  in the context of black holes in~\cite{Gurlebeck:2015xpa,Poisson:2020vap,Poisson:2021yau,Riva:2023rcm,Iteanu:2024dvx,Combaluzier-Szteinsznaider:2024sgb,DeLuca:2023mio,Kehagias:2024rtz}. For neutron stars, nonlinear tidal effects have been first explored  in~\cite{Yu:2022fzw} (see also~\cite{Schenk:2001zm,Weinberg_2012}), which suggested that nonlinearities can induce sizable corrections to the GW phase relative to linear predictions, although that conclusion relied on a mode decomposition of the tidal response and a Newtonian description of the stellar interior with a simplified EoS. A relativistic generalization was more recently discussed in~\cite{Pitre:2025qdf} (see also~\cite{Poisson:2020vap}). Our work goes beyond these studies in several ways:  (1)~Unlike~\cite{Yu:2022fzw}, we do  not rely on a mode representation of the tidal
deformation, but work in a fully general-relativistic framework (as in~\cite{Pitre:2025qdf}), solving the Einstein equations up to quadratic order in perturbation theory; (2)~In contrast to~\cite{Pitre:2025qdf}, we define the nonlinear Love numbers within the point-particle effective field theory~(EFT), which provides an alternative coordinate-independent and unambiguous way to characterize tidal responses; (3)~We numerically solve the nonlinear perturbation equations in the stellar interior for realistic EoS and compute the leading correction to the GW phase. This allows us to estimate the impact of quadratic Love numbers on the gravitational waveform and compare it to other subleading effects.

\vspace{0.1in}
\noindent
\textit{Outline:}
The paper is organized as follows. In Section~\ref{sec:EFT}, we introduce the worldline EFT and compute the induced static response metric (see Eq.~\eqref{eq:deltagttEFT}) at second order in perturbation theory and leading order in the gradient expansion. In Section~\ref{sec:StellarP},  we  derive the equations governing the dynamics of the parity-even fields up to second order in perturbations in a fully general-relativistic setup. 
As anticipated, we focus on the leading-order quadrupolar static response. 
In Section~\ref{QLN}, we match the EFT result~\eqref{eq:deltagttEFT} in the static limit to
the full solution obtained from numerical integration of the perturbation equations inside the star, and compute the quadratic Love numbers for various realistic, nuclear-physics motivated EoS. The impact on the GW waveform is discussed in Section~\ref{sec:waveform}. Some complementary and more technical results are collected in Appendices~\ref{app:FDres}, \ref{angIntApp} and~\ref{2ndOrderApp}. 

\vspace{0.1in}
\noindent
\textit{Conventions:}
We use the mostly-plus signature for the metric, $(-, +, +, +)$, and work in natural units, $\hbar = c = 1$. We denote the reduced Planck mass by $ \Mpl=  1/\sqrt{8 \pi G}$,
and use the curvature convention ${R^\rho}_{\sigma\mu\nu}=\partial_\mu\Gamma^\rho_{\nu\sigma}+\dots$ and $R_{\mu\nu}={R^\rho}_{\mu\rho\nu}$.
Our convention for the decomposition in spherical harmonics and the Fourier transform is 
$\psi(t,r, \theta,\varphi)= \sum_{\ell,m}\int\frac{\D\omega}{2\pi} \e^{-i\omega t}  \psi^{\ell m}(\omega,r)Y_{\ell m}( \theta,\varphi)$.  For simplicity, we will often omit the arguments on ${\psi^{\ell m}}$ altogether, relying on the context to discriminate between the different meanings.  
Throughout we use capital Latin letters $A, B, C,\cdots,$ to denote angular indices on the two-dimensional   sphere $S^2$.


\section{Worldline Effective Field Theory}
\label{sec:EFT}

\subsection{The action}

A robust way to define the tidal response of a compact object is through the worldline EFT  framework~\cite{Goldberger:2004jt, Goldberger:2007hy, Porto:2016pyg}. 
In this approach, a compact object of radius $\Rs$ is viewed from a sufficiently large distance $r \gg \Rs$, so that it can be treated, at leading order, as a point particle. Finite-size effects are then systematically incorporated by adding all possible non-minimal, non-redundant operators consistent with the symmetries of the long-distance physics.

In particular,  the Einstein--Hilbert action describes the bulk gravitational dynamics,
\be	S_{\rm EH}  = \int \D^4 x \sqrt{-g}
	\frac{\Mpl^2}{2} R  \; ,
\ee
while, in the point-particle approximation, the dynamics of  the compact object is given by the worldline action 
\begin{align}
	S_{\rm pp}  = -m \int \D \tau = 
	-m \int \D \sigma 
	\sqrt{-g_{\mu\nu}(X)\frac{\D X^\mu}{\D \sigma}\frac{\D X^\nu}{\D \sigma}} \; ,
	\label{eq:Spp}
\end{align}
where $\tau$ is the proper time along the particle's worldline,  $m$ is its mass, $X^\mu(\sigma)$ denotes its trajectory, and $\sigma$ is an affine parameter that parametrizes the worldline.

To account for finite-size effects, we include in the action all possible non-minimal operators localized on the point particle's worldline that are consistent with  diffeomorphism invariance and worldline reparametrization invariance, organized by the number of fields and derivatives. We focus here on the static response, excluding operators with derivatives in the proper time~\cite{Goldberger:2005cd, Goldberger:2020fot,Saketh:2024juq}, and restrict our attention to the leading linear and quadratic contributions, i.e.~those describing the response of a quadrupolar deformation to a quadrupolar tidal field ($\ell=2$), in the parity-even sector.\footnote{In the PN expansion, the parity-odd response is suppressed by a relative-velocity factor compared to the electric-type tidal response~\cite{Yagi:2013sva,Henry:2019xhg}.} In this case, the tidal action can be written as~\cite{Goldberger:2004jt,Damour:2009vw,Bern:2020uwk,Riva:2023rcm,Iteanu:2024dvx}
\begin{equation}
	S_{\rm T} = \int \D \tau \left(
	\lambda_2 
	E_{\mu}{}^{\nu} E_{\nu}{}^{\mu}
	+ \lambda_{222}
	E_{\mu}{}^{\rho} E_{\rho}{}^{\nu}E_{\nu}{}^{\mu} +\dots
	\right) \; ,
	\label{eq:Stidal}
\end{equation}
where $E_{\mu\nu}$ is the electric component of the Weyl tensor,
\begin{equation}
	E_{\mu\nu} \equiv C_{\mu\rho\nu\sigma}U^\rho U^\sigma \; ,
\end{equation}
and $U^\mu = \D X^\mu/\D \tau$ is the four-velocity of the point particle,  normalized as $g_{\mu\nu} U^\mu U^\nu = -1$. Moreover, $\lambda_2$ and $\lambda_{222}$ are Wilson coefficients that encode the object's linear and quadratic quadrupolar responses, respectively. The $\lambda_2$ is the usual linear Love number while $\lambda_{222}$ is the quadratic Love number.

{It is convenient to express these coefficients in terms of the following dimensionless combinations,
\begin{equation}
k_2 \equiv 6 \lambda_2 \frac{G}{\Rs^5} \;, \qquad
p_2 \equiv - 9 \lambda_{222} \frac{ G^2 m}{\Rs^8} \;,
\label{dimscale}
\end{equation}
where the numerical prefactors are chosen to match standard conventions in the literature~\cite{Hinderer:2007mb,Pitre:2025qdf}\footnote{Note that the sign of $p_2$ in~\cite{Pitre:2025qdf} is opposite to the one used in~\cite{Poisson:2020vap}.} and the scaling factors follow from dimensional analysis.}\footnote{By restoring factors of the speed of light so that the limit $c \to \infty$ reproduces the Newtonian result~\cite{Henry:2019xhg}, the action of the body becomes
\begin{equation}
	\int \D \tau \left( - m c^2+
	c^4\lambda_2 
	 E_{\mu}{}^{\nu} E_{\nu}{}^{\mu}
	+ c^6\lambda_{222}
	E_{\mu}{}^{\rho} E_{\rho}{}^{\nu}E_{\nu}{}^{\mu} +\dots
	\right) \; .
    \label{scalingac}
\end{equation}
The combination $c^2 E_{\mu\nu}$ has dimensions $T^{-2}$, where $T$ is a characteristic timescale. Taking $T$ to be the  dynamical timescale of the object, $T \sim (\Rs^3 / G m)^{1/2}$, and requiring Eq.~\eqref{scalingac} to have the dimension of an action, one directly obtains the scaling behavior shown in Eq.~\eqref{dimscale}.
}

For the following computation, we will start from the  EFT action
\begin{equation}
	S_{\rm EFT} = S_{\rm EH} + S_{\rm pp} + S_{\rm T} \; .
	\label{eq:SEFT}
\end{equation}
Note that one can straightforwardly extend the above action to include higher-derivative operators and the odd sector. See e.g.~\cite{Bern:2020uwk,Haddad:2020que,Riva:2023rcm,Iteanu:2024dvx,Combaluzier-Szteinsznaider:2024sgb} for more details.

\subsection{Quadrupolar metric perturbation}
\label{sec:EFTquadrupole}

The goal of this section is to compute the metric perturbation from the worldline action~\eqref{eq:SEFT} and to match it with the result obtained in Section~\ref{sec:StellarP} in the long-distance limit, $r \gg \Rs \ge r_s$, with $r_s \equiv 2 G m$ being the Schwarzschild radius of the object. This matching procedure will allow us to determine the values of the Love numbers $\lambda_2$ and $\lambda_{222}$.

Following~\cite{Riva:2023rcm,Iteanu:2024dvx}, we employ the background-field method~\cite{DeWitt:1967ub,tHooft:1974toh,Abbott:1980hw}, and define the canonically normalized metric perturbation,
\begin{equation}
h_{\mu \nu} \equiv 2 \Mpl ( g_{\mu\nu} - \bar{g}_{\mu\nu} ) \; ,
\end{equation}
where the background metric is written as $\bar{g}_{\mu\nu} = \eta_{\mu\nu} + H_{\mu\nu}$, with $H_{\mu \nu}$ representing the external tidal field. The latter satisfies the vacuum Einstein equations and is chosen to vanish on the worldline, $H_{\mu\nu}(X) = 0$.

With the above definition, the  metric perturbation $h_{\mu\nu}$ we wish to solve  for can be written as
\begin{equation}
\delta g^{\rm EFT}_{\mu \nu} =
H_{\mu\nu} + \frac{\langle h_{\mu\nu}\rangle}{2 \Mpl} \; ,
\end{equation}
where $\langle h_{\mu\nu}\rangle$ denotes the one-point function generated by the interaction between the external tidal field and the source described by the action~\eqref{eq:SEFT}.
Formally, this expectation value can be computed through the path integral,
\begin{equation}
\langle h_{\mu\nu}\rangle
= \int \mathcal{D}[h] \, h_{\mu\nu} 
\e^{i(S_{\rm EFT} + S_{\rm GF})} \; ,
\label{eq:Pathh}
\end{equation}
up to an overall normalization. Here $S_{\rm GF}$ denotes the standard gauge-fixing term introduced by the Faddeev--Popov procedure. To preserve covariance of the result with respect to the background metric, we adopt the harmonic (de Donder) gauge in the background-field formulation,
\be
 S_{\text{GF}}  = -\int \dd^4 x \sqrt{-\bar{g}} \bar{g}^{\mu\nu}
		\bar{\Gamma}_\mu \bar{\Gamma}_\nu \, , 
        \ee
        with
        \be
 \bar{\Gamma}_\mu  \equiv \bar{g}^{\alpha\beta}\bar{\nabla}_{\alpha} h_{\beta\mu} - 
		\frac{\bar{g}^{\alpha\beta}}{2}\bar{\nabla}_\mu h_{\alpha\beta} \; ,
		\label{eq:GFaction}
\ee
where $\bar{\nabla}_\mu$ is the covariant derivative compatible with the background metric $\bar{g}_{\mu\nu}$.

\begin{figure}[t]
\begin{center}
	\includegraphics[width=0.47\textwidth]{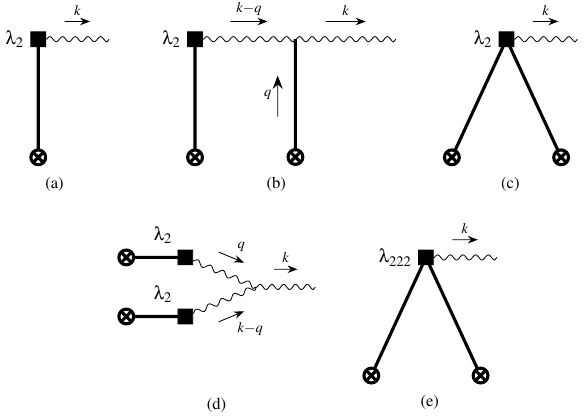}
\end{center}
\caption{Feynman diagrams needed to match $\lambda_2$ and  $\lambda_{222}$ at leading order in powers of $r_s$. Diagram (a) corresponds to the linear response, while (b) and (c) represent nonlinear corrections to it. Diagram (d) describes the nonlinear interaction between two linear responses, while diagram (e) represents the genuine nonlinear response proportional to the nonlinear quadrupole coefficient $\lambda_{222}$.}
\label{fig:FeynD}
\end{figure}

In practice, we compute the above quantity diagrammatically by considering all connected tree-level Feynman diagrams with one $h_{\mu\nu}$ external leg. Ignoring graviton loop diagrams ensures that the final result is entirely classical~\cite{Goldberger:2004jt}. In particular, we introduce the following diagrammatic conventions:
\begin{align*}
	\raisebox{3pt}{\gravconv} & \equiv \text{metric perturbation } h \, , \\
	\raisebox{3pt}{\TidalHconv} & \equiv \text{tidal field } H \, ,\\
	\raisebox{3pt}{\MassSc} & \equiv \text{point-particle source} \; , \\
	\raisebox{3pt}{\LinT} & \equiv \text{tidal source} \; .
\end{align*}

To determine the value of the nonlinear coefficient $\lambda_{222}$, we evaluate the diagrams shown in Fig.~\ref{fig:FeynD}. In particular, we consider only diagrams that contribute to the induced second-order quadrupole, that is, the $\ell=2$ moment at the nonlinear level. To simplify the computation, we evaluate the diagrams in the rest frame of the object, placed at the origin of the coordinate system. We therefore choose $\sigma = t$ as the affine parameter, so that
\begin{equation}
	X^\mu(t) = (t, \mathbf{0}) \; ,
	\qquad
	\frac{\D X^\mu(t)}{\D t} = (1, \mathbf{0}) \; .
\end{equation}
Moreover, as explained in~\cite{Riva:2023rcm,Iteanu:2024dvx}, since we are working in the background gauge~\eqref{eq:GFaction}, we are free to choose any convenient gauge for the external tidal field ${H}_{\mu\nu}$. For later convenience then, we work with the external tidal field written in the Regge--Wheeler~(RW) gauge, defined perturbatively order by order around flat spacetime. For example, the leading-order tidal field is given by
\begin{equation}
	H_{tt} = \sum_m \mathscr{E}_{2, m}r^2 Y_2^m(\theta, \varphi)\; , 
	\quad 
	H_{ij} = \delta_{ij}\sum_m \mathscr{E}_{2, m} r^2 Y_2^m(\theta, \varphi) \; ,
\end{equation}
where $\delta_{ij}$ is the flat Euclidean metric. Here, $\mathscr{E}_{2, m}$ is the amplitude of the external tidal field and has dimensions of an inverse length squared. Explicit expressions for the second-order tidal field can be found in~\cite{Iteanu:2024dvx}.

The evaluation of the diagrams shown in Fig.~\ref{fig:FeynD} follows straightforwardly from the corresponding Feynman rules. Their explicit expressions are not particularly illuminating; we therefore report the relevant result for each diagram in Appendix~\ref{app:FDres}.

As a consistency check, we verified that the computed one-point function $\langle h_{\mu\nu} \rangle$ satisfies the gauge condition
\begin{equation}
	\bar{\Gamma}_\mu = 0 \; .
    \label{gaugec}
\end{equation}
Note that Fig.~\ref{fig:FeynD}(c) represents a nonlinear response arising from the linear operator $\lambda_2$ in Eq.~\eqref{eq:Stidal}. This contribution appears due to the intrinsic nonlinearity of general relativity and to ensure that the EFT effective action~\eqref{eq:SEFT} is invariant under a generic diffeomorphism. Indeed, the main role of diagram~\ref{fig:FeynD}(c) is to guarantee the covariance of the final result and that the metric perturbation satisfies the gauge condition above.

The resulting metric perturbation will be matched to the one obtained using stellar perturbation theory in Section~\ref{sec:StellarP}. There, both the calculation and the final result are presented in the RW gauge~\cite{Regge:1957td}, which for polar perturbations is defined by the conditions $g_{tA}=g_{rA}=0$ and the absence of angular off-diagonal components; see details in the next section. Consequently, we must perform a second-order gauge transformation from the coordinates 
$x^\mu$
 satisfying the gauge condition~\eqref{gaugec} to the RW coordinates 
$x^\mu_{\rm RW}$ employed in Section~\ref{sec:StellarP}.
Explicitly, we transform the full metric $g_{\mu\nu}(x) = \bar{g}_{\mu\nu}(x) + h_{\mu\nu}(x)/(2\Mpl)$ as usual
\begin{equation}
	g^{\rm RW}_{\mu\nu}(x_{\rm RW}) = \frac{\partial x^\rho}{\partial x_{\rm RW}^\mu}\frac{\partial x^\sigma}{\partial x_{\rm RW}^\nu} g_{\rho\sigma}(x) \; .
\label{gaugetran}
\end{equation}
We consider a perturbative gauge transformation $x^\mu_{\rm RW} = x^\mu + \xi^\mu$ and we recall that the tidal background metric is already expressed in RW coordinates. Expanding Eq.~\eqref{gaugetran} for small $\xi^\mu$, and focusing on the $tt$ component of the metric in the static limit, the relevant transformation  is
\begin{equation}
	g^{\rm RW}_{tt} =  
	 \left(1 - \xi^i \partial_i - \frac12    	
	\xi^i \xi^j \partial_i \partial_{j} \right) \bar{g}_{tt} + (1- \xi^i \partial_i) \frac{h_{tt} }{2\Mpl}  \; .
\end{equation}
Higher-order terms in $\xi$  contribute either at higher order in $r_s$ or at higher order in ${\cal E}$ and can therefore be  neglected.
Focusing on the $\ell=2$ multipole and imposing that the final metric satisfies the RW gauge condition at the relevant order in $r_s$ and ${\cal E}$, we find that the  gauge transformation is
\begin{align}
\xi_{(\ell=2,m)}^i = & \ \frac{r_s}{2}n^i
	 +\sum_m \mathscr{E}_{2, m} \left[ \frac{r_s r^2}{2}
	\big(
	n^i + r\,\gamma^{ij}\partial_j
	\big)  
    \notag \right. \\
	& + \left.   \frac{2 G \lambda_2 }{r^2}
	\big(
	2n^i + r\,\gamma^{ij}\partial_j
	\big) \right] Y_2^m(\theta, \varphi)
	\; ,
\end{align}
where $n^i = x^i/r$ is the normal radial direction and  $\gamma_{ij} = \delta_{ij}-n_i n_j$ is the metric of the two-sphere written in Cartesian coordinates. The first term is the standard gauge transformation for the Schwarzschild metric from harmonic to Schwarzschild coordinates, while the first term inside the bracket corresponds to the transformation of the $r_s$ contribution to the tidal field computed in~\cite{Riva:2023rcm, Iteanu:2024dvx}. The last term gives the needed gauge transformation of the linear response contribution from Fig.~\ref{fig:FeynD}(a).

At this order, the $tt$ component of the nonlinear response given by Fig.~\ref{fig:FeynD}(e) is gauge-invariant. Altogether, at leading order in $r_s$, the full $\ell=2$ part of the $tt$ component of the EFT metric computation is
\begin{widetext}
\begin{equation}
	\delta g^{{\rm EFT}(\ell=2, m)}_{tt} (r) =  
	\mathscr{E}_{2, m} \left( r^2  + 12 \frac{G \lambda_2}{r^3} \right)
	-\sum_{m_1 m_2}I^{m m_1 m_2}_{2 2 2} \mathscr{E}_{2, m_1} \mathscr{E}_{2, m_2} \left[\frac{1}{2}r^4
	  + \left(
	12\frac{G \lambda_2}{r}
	+\frac{63}{2}\frac{G \lambda_{222}}{ r^3}
	+72\frac{G^2\lambda_2^2}{ r^6}
	\right) \right] \; ,
\label{eq:deltagttEFT}
\end{equation}
\end{widetext}
where we have introduced the angular integral
\begin{equation}
I^{m m_1 m_2}_{\ell \ell_1 \ell_2} \equiv \int \D \Omega \, Y^{*}_{\ell m} Y_{\ell_1 m_1} Y_{\ell_2 m_2} \;.    
\end{equation}
The first term contains the standard linear tidal contribution as well as the static linear response, which comes entirely from diagram~\ref{fig:FeynD}(a).
The second term encodes  the nonlinear external tidal field---scaling as $r^{4}$---and the nonlinear response. In particular, the contribution proportional to $\lambda_{2}$ gives the leading nonlinear correction to the linear response and originates from diagrams~\ref{fig:FeynD}(b) and~\ref{fig:FeynD}(c). 
Diagram~\ref{fig:FeynD}(d) generates the term quadratic in the linear response, proportional to $\lambda_{2}^{2}$, while diagram~\ref{fig:FeynD}(e) produces the genuinely nonlinear response proportional to $\lambda_{222}$.

In Section~\ref{QLN} we will match  Eq.~\eqref{eq:deltagttEFT} to the full metric solution obtained from solving the stellar perturbation equations for $\ell=2$ in general relativity. Before proceeding, it is worth making some important remarks.  Note that the expression~\eqref{eq:deltagttEFT} does not contain all the terms  that contribute to the shown falloffs---in particular the  $r^2$ tidal profile and the $r^{-3}$ response. Equation \eqref{eq:deltagttEFT} was obtained from a tree-level worldline calculation and does not yet include gravitational effects. There are two types of gravitational corrections---resulting from background nonlinearities---that one would need to compute in order to obtain a full matching at subleading order in $r_s$. 
The first one arises from a $2^{\rm nd}$-order $r_s$-correction to the  response term $\mathscr{E}^2\lambda_2/r$ in Eq.~\eqref{eq:deltagttEFT}. This  yields a contribution of the form $\mathscr{E}^2\lambda_2 r_s^2/r^3$, which overlaps with the quadratic response  $\mathscr{E}^2\lambda_{222}/r^3$---although with a different scaling in  $r_s$,  resulting in a relative factor $(r_s/R_*)^3$.
In addition, there are subleading corrections to the tidal field obtained from higher-order diagrams computed solely from $S_{\rm EH}$ and $S_{\rm pp}$ in the EFT action~\eqref{eq:SEFT} (i.e., without  insertions from $S_{\rm T}$).
An example is the $5^{\rm th}$-order (respectively, $7^{\rm th}$-order) $r_s$-correction to the linear (quadratic) tidal field $\mathcal{E}r^2$ ($\mathcal{E}^2 r^4$), which would generate additional $1/r^3$ contributions  to~\eqref{eq:deltagttEFT}. However, such terms can be shown to be absent at both linear and  nonlinear order~\cite{Parra-Martinez:2025bcu, inprogress}. 
Moreover, there is a $2^{\rm nd}$-order $r_s$-correction to the quadratic tidal field $\mathcal{E}^2 r^4$, which scales like $\mathscr{E}^2r_s^2r^2$. If not computed, this could introduce an ambiguity in the matching with the full general-relativistic tidal field solution. Note that such a term can be  formally reabsorbed into a redefinition of the tidal amplitude, i.e.~$\mathscr{E} \rightarrow \mathscr{E} + \# \mathscr{E}^2 r_s^2$, which is tantamount to  correcting Eq.~\eqref{eq:deltagttEFT} by a term of the form $\sim \mathscr{E}^2\lambda_2 r_s^2/r^3$. This has  the same scaling as the nonlinear response $\mathscr{E}^2\lambda_{222}/r^3$ but, like the $2^{\rm nd}$-order $r_s$-correction term mentioned above,  carries a relative $(r_s/R_*)^3$ factor. Although these effects should in principle be included in order to isolate the true nonlinear response, in practice we will simply  ignore them in what follows. They in fact contribute to the value of the $p_2$ coefficient through a correction proportional to $C^3 k_2(C) $, where $C$ is the compactness defined in Eq.~\eqref{eq:defC}, which is  small for the typical neutron-star compactness values of interest,\footnote{For example, for $0.1 \lesssim C \lesssim 0.2$ and $k_2 \lesssim 0.1$ (see, e.g.,~Fig.~\ref{fig:MR}) one obtains  $C^3 k_2 \lesssim 10^{-3}$.} and vanish in the black hole limit, $k_2^{\rm BH} =0$ \cite{Damour:2009vw,Binnington:2009bb}.
Since we are primarily interested in estimating nonlinear tidal effects for realistic neutron stars, we will ignore these corrections in the present work. A complete analysis will be provided elsewhere.

We now turn to the computation of the metric in the full theory, which will be matched to the expression above in order to extract the value of static Love numbers $\lambda_2$ and $\lambda_{222}$.


\section{Stellar perturbation theory} 
\label{sec:StellarP} 

In this section we derive the  equations governing the dynamics of the perturbations of a relativistic star, up to second order in the perturbative expansion. We will assume the unperturbed star to be described by a spherically-symmetric background metric $\bar{g}_{\mu\nu}$, with line element 
\begin{equation}
\D s^2= \bar{g}_{\mu\nu} \D x^\mu\D x^\nu = - \e^{\Phi(r)}\D t^2 + \e^{\Psi(r)}\D r^2 + r^2 \D \Omega_{S^2}^2 ,
\end{equation}
where $\Phi(r)$ and $\Psi(r)$ are functions of the radial coordinate $r$ only, and $\D \Omega_{S^2}^2 \equiv \D \theta^2+ \sin^2\theta \D \varphi^2$ denotes the line element on the two-sphere. 

We further assume the interior of the  star to be described by the energy-momentum tensor of a perfect fluid,
\begin{equation}
T_{\mu\nu} = (\rho+p)u_\mu u_\nu+ p g_{\mu\nu} ,
\end{equation}
where $\rho$ denotes   the energy density, $p$ the pressure, and $u_\mu$ the fluid four-velocity.
At the background level, we shall denote with $\bar T_{\mu\nu} = (\bar\rho+\bar p)\bar{u}_\mu \bar{u}_\nu+ \bar{p} \bar{g}_{\mu\nu}$ the unperturbed energy-momentum tensor of the star.

For convenience, in this section we set $G=1$.\footnote{We  restore  explicit dependence of $G$  in Section~\ref{QLN}.} Then, from the Einstein equations, 
\begin{equation}
G_{\mu\nu} = 8 \pi T_{\mu\nu} \, ,
\end{equation}
one finds that, at the background level, $\bar{g}_{\mu\nu}$ and the unperturbed density $(\bar \rho)$, pressure  $(\bar p)$ and fluid velocity $(\bar{u}_{\mu})$ in the matter sector are related via the Tolman--Oppenheimer--Volkoff (TOV) equations \cite{Tolman:1939jz,Oppenheimer:1939ne},
\begin{equation}
\Mm'(r)=4\pi r^2\bar \rho(r) \; ,
\label{TOV1}
\end{equation}
\begin{equation}
\Phi '(r)= 2\frac{ \Mm(r)+4 \pi  r^3 \bar p(r)}{r[r-2  \Mm(r)]} \; ,
\label{TOV2}
\end{equation}
\begin{equation}
\bar p'(r) =  -[\bar p(r)+\bar \rho (r)]\frac{\left[\Mm(r)+4 \pi  r^3 \bar p(r)\right]}{r [r-2 \Mm(r)]} \; ,
\label{TOV3}
\end{equation}
where we have defined $\Mm(r)$ by $\e^{-\Psi(r)} \equiv 1-2\Mm(r)/r$. Since the background is described by four quantities, $\left(\bar p(r),\bar\rho(r),\Mm(r),\Phi(r) \right)$, closing the system requires specifying an EoS. We will consider several representative EoS when presenting our results in Section~\ref{sec:numres}. In addition, given $u_\mu u^\mu = -1$, the  unperturbed fluid velocity is found to be
\begin{equation}
\bar{u}^\mu = \begin{pmatrix}
\e^{-\Phi(r)/2}, & 0, & 0, & 0
\end{pmatrix} .
\end{equation}

Outside the star, the metric reduces to the Schwarzschild solution, $\e^\Phi = \e^{-\Psi}= 1 - \frac{2 m}{r}$, with 
\be
m \equiv \Mm(\Rs) \;
\ee
denoting the total mass of the star.

We shall then expand both the metric and the matter sector in perturbations as $g_{\mu\nu}=\bar{g}_{\mu\nu}+ \delta g_{\mu\nu}$ and $T_{\mu\nu}=\bar{T}_{\mu\nu}+ \delta T_{\mu\nu}$. For simplicity, we will focus here on the parity-\textit{even} (or, \textit{polar}) sector. 

The most general parametrizations for  the fluid four-velocity and the even metric fluctuation are~\cite{Regge:1957td}:
\begin{equation}
\delta u^\mu = \frac{\e^{\Phi/2}}{4\pi(\bar{\rho}+\bar p)}
\begin{pmatrix}
\delta u^0, & \e^{-\Psi}R, & \frac{1}{r^2}\gamma^{AB}\nabla_B V
\end{pmatrix} ,
\end{equation}
\begin{equation}
\delta g_{\mu\nu} =
\begin{pmatrix}
\e^\Phi H_0 & H_1 & \nabla_A {\mathcal H}_0 \\
* & \e^\Psi H_2 & \nabla_A {\mathcal H}_1 \\
*  & * & \,\,\,\, r^2 { K}\gamma_{AB} + r^2(\nabla_A \nabla_B - \frac{1}{2} \gamma_{AB}\nabla^C\nabla_C) G
\end{pmatrix},
\label{RWgeven}
\end{equation}
where the  entries denoted by an asterisk are equal to the corresponding entries across the diagonal, because $\delta g_{\mu\nu}$ is symmetric. The quantities $H_0$, $H_1$, $H_2$, ${\mathcal H}_0$, ${\mathcal H}_1$, $K$, $G$, $\delta u^0$, $R$ and $V$ are functions of the spacetime coordinates $(t,r,\theta,\varphi)$, $\gamma_{A B}$ is the two-dimensional metric on $S^2$, $\gamma_{A B} {\rm d}x^A {\rm d}x^B \equiv \D \Omega_{S^2}^2    = {\rm d}\theta^2 + \sin^2\theta\, {\rm d}\varphi^2$, and 
 $\nabla_A$ is the  covariant derivative on $S^2$.\footnote{We are following here the same notation  and conventions of~\cite{Iteanu:2024dvx}, to which we refer for details.} 
The normalization condition of the four-velocity, $u^\mu u_\mu=-1$, fixes $\delta u^0$ in terms of the other fluctuations. Up to second order in perturbation theory, one finds:
\begin{multline}
\delta u^0 = 2 \pi  \e^{-\Phi} (\bar \rho + \bar p) H_0
+ \frac{3\pi}{2}  \e^{-\Phi} (\bar \rho + \bar p) H_0^2
\\
+ \e^{-\Psi} \left[ \e^{-\Phi} H_1 R 
+\frac{r^2 R^2+\e^{\Psi} \nabla_A V \nabla^A V }{8\pi  r^2 (\bar p+\bar \rho )}\right].
\end{multline}
In the following, we  work in the RW gauge~\cite{Regge:1957td}, defined such that ${\mathcal H}_0={\mathcal H}_1=G=0$.

Furthermore, we move to frequency space and expand all scalar functions of the spatial coordinates in terms of spherical harmonics:
\begin{equation}
    X(t,\vec x) = \int \frac{\D \omega}{2\pi} \sum_{\ell m} \e^{-i\omega t} Y_{\ell m}(\theta,\varphi) X^{\ell m}(\omega,r) .
\end{equation}

\subsection{Linear perturbations and zero-frequency limit}

Let us first expand the Einstein equations at linear order in perturbation theory~\cite{Chandrasekhar:1991fi,Ipser:1991ind,1991RSPSA.433..423C, Kojima:1992ie, 1967ApJ...149..591T,Detweiler:1985zz}. Their solutions  will later enter the second-order equations as source terms, discussed further below.

Since linear perturbations of different frequencies and angular numbers decouple due to the symmetries of the background, we will drop the corresponding labels to ease the notation. 

At linear order, the Einstein equations are given by 
\begin{equation}
0=\Ef_{\mu\nu}\equiv {}^{(1)}\!G_{\mu\nu}- 8 \pi \, {}^{(1)}\!T_{\mu\nu} \;,
\end{equation}
where the left superscript indicates that each quantity has been expanded in perturbations up to the specified order.
Since $\Ef_{\mu\nu}$ is a rank-two tensor, it admits a decomposition into tensor spherical harmonics in the same basis as the metric perturbations. Projecting out the angular dependence using the identities from App.~\ref{angIntApp}, and using a prime to denote  the radial derivative, the components of $\Ef_{\mu \nu}$ are given by 
\begin{widetext}
\begin{align}
    \Ef_{tt}  = & \ \frac{\e^{\Phi }}{2 r^2}  \big\{  2 \left[(r-2 \Mm) (\Htwof'-r \, \Kf'')+ \left(5 \Mm+4 \pi  r^3 \bar \rho-3 r\right) \Kf' -8 \pi  r^2 \deltaf  \rho
   \right] \nonumber \\
   &  + \left(\ell(\ell+1)-16 \pi  r^2 \bar\rho +2\right)\Htwof + \left(\ell(\ell+1)-2\right) \Kf \big\} Y_{\ell m}(\theta, \varphi) \;, \label{Ett}\\
\Ef_{tr}  = & \  \frac{1}{2 r^2} \left\{\Honef \ell(\ell+1)+2 r \left[-i \omega  ( \Htwof - r  \Kf') -\frac{i \omega   \left(3 \Mm+4 \pi  r^3 \bar p-r\right) \Kf }{r-2 \Mm}+2 r
   \e^{\Phi } \Rf\right] \right\} Y_{\ell m}(\theta, \varphi) \;, \\
    \Ef_{rr}  = & \ \frac{\e^{-\Phi }}{2 r (r-2 \Mm)} \big\{ -\e^{\Phi } \big[2 (r-2 \Mm) \Hzerf' + 2 \left(1+ 8 \pi  r^2 \bar p \right) \Htwof  + 2\left(\Mm-4 \pi  r^3
   \bar p-r\right) \Kf' +16 \pi  r^2 \deltaf p \nonumber \\
   &  -\ell(\ell+1) \Hzerf +  (\ell(\ell+1)-2 ) \Kf \big] +2 \omega  \big(r^2 \omega 
   \Kf-2 i (r-2 \Mm) \Honef  \big) \big\} Y_{\ell m}(\theta, \varphi) \;, \\
\gamma^{AB} \Ef_{AB}   = & \ \frac{\e^{-\Phi }}{2}  \big\{ \e^{\Phi } \big[  \ell(\ell+1) \Hzerf - \left(\ell(\ell+1)+32 \pi  r^2 \bar p\right) \Htwof -2  \left(r+\Mm+4 \pi 
   r^3 (2 \bar p- \bar \rho ) \right) \Hzerf' -2  \left(r-\Mm+4 \pi  r^3 \bar p\right) \Htwof'  
   \nonumber \\
&     
-2 r (r-2 \Mm) (\Hzerf''- \Kf'') +4 
   \left(r -\Mm+2 \pi  r^3 ( \bar p-\bar \rho )\right) \Kf' -32 \pi  r^2 \deltaf p\big] \nonumber \\  
   & +2 r \omega  \left[r \omega 
   \left(\Htwof+\Kf\right)-2 i (r-2 \Mm) \Honef'\right] 
   +4 i \omega  \Honef \left(\Mm+4 \pi  r^3 \bar \rho -r\right)\big\} Y_{\ell m}(\theta, \varphi) \;, \\
    \Ef_{tA} =  & \ \frac{1}{2 r^2} \left\{  r \left[ (r-2 \Mm) \Honef'+4 \pi  r^2 ( \bar p- \bar \rho ) \Honef  +i r \omega  \Kf+4 r \e^{\Phi } \Vf\right]+2
    \Mm \Honef +i r^2 \omega  \Htwof \right\} \nabla_A Y_{\ell m}(\theta, \varphi) \;, \\
    \Ef_{rA}  = & \ \frac{1}{2}  \left[ \frac{  (3 \Mm -r +4 \pi  r^3 \bar p) \Hzerf - (\Mm-r+4 \pi  r^3 \bar p) \Htwof  }{r (r-2 \Mm)}+i
   \omega  \e^{-\Phi } \Honef  +\Hzerf'-\Kf'\right] \nabla_A Y_{\ell m}(\theta, \varphi) \;, \\
    [\Ef_{AB}]_{\rm{STF}}  = & \ \frac{1}{2} \left(\Hzerf-\Htwof\right)  \left( \nabla_A\nabla_B  -\frac{1}{2} \gamma_{AB} \nabla_C\nabla^C \right) Y_{\ell m}(\theta, \varphi)   \;,
    \label{EAB}
\end{align}
\end{widetext}
where $[\cdots]_{\rm STF}$ in~\eqref{EAB}  denotes the symmetric trace-free combination.  Note that we have used the background TOV equations, Eqs.~\eqref{TOV1}-\eqref{TOV3}, to simplify these expressions. This is a system of seven equations for eight perturbation variables,
$\Hzerf$, $\Honef$, $\Htwof$, $\Kf$, $\Rf$, $\Vf$, $\deltaf \rho$, and $\deltaf p$. In general, to close the system we need to specify the EoS.

From Eq.~\eqref{EAB}, $[\Ef_{AB}]_{\rm STF}=0$ yields the constraint $\Htwof = \Hzerf$. The equations  $\Ef_{tr} = 0$ and $\Ef_{tA}=0$ allow one, in general, to express $\Rf$ and $\Vf$  in terms of $\Hzerf, \Honef, \Kf$, and their derivatives. Additionally, $\deltaf p$ (and therefore $\deltaf \rho$ via the EoS) can be written in terms of the same variables   by using $\Ef_{rr}=0$. The remaining quantities, $\Hzerf$,  $ \Honef$ and $\Kf$, are then determined by solving the set of equations $\Ef_{tt} =0$, $ \Ef_{rA} =0 $, and $ \gamma^{AB}\Ef_{AB} = 0$~\cite{Chandrasekhar:1991fi,Ipser:1991ind,1991RSPSA.433..423C,Kojima:1992ie,Ipser:1991ind}. 
So far, we have kept the frequency generic; as such, Eqs.~\eqref{Ett}-\eqref{EAB} are fully general.
This is convenient for addressing a potential issue in finding the static solutions, as we now discuss.

Since we are interested in studying the nonlinear static response, we would like to take the zero-frequency limit $\omega\rightarrow0$.
This limit can, however, be subtle. In fact, setting $\omega=0$ directly in Eqs.~\eqref{Ett}-\eqref{EAB} leaves  the variables  $\Vf$, $\Rf$, $\Honef$  apparently unconstrained: as mentioned, one can use $\Ef_{tr}=0$ and $\Ef_{tA}=0$ to express $\Vf$ and $\Rf$ in terms of $\Honef$, but $\Honef$ disappears from all the remaining equations when $\omega$ vanishes exactly (in particular, we can no longer use e.g.~$ \Ef_{rA} =0 $ to solve for $\Honef$ in terms of $\Htwof$, $\Hzerf$ and $\Kf$).
Although this is not an issue in principle for studying the linear static response---where $\Honef$ is not needed explicitly to solve for $\Hzerf$~\cite{Hinderer:2007mb}---it becomes problematic at second order, as this would leave some ambiguity in the source and in the second-order solution. The issue is resolved by taking the  $\omega\rightarrow0$ limit carefully~\cite{Pani:2018inf}: after solving the constraints at $\omega\neq0$, one finds that the frequency appears only quadratically in the remaining equations for $\Kf$ and $\Hzerf$ (see, e.g.,~\cite{Ipser:1991ind,Kojima:1992ie}). This implies that, for these variables, the solutions at orders $\mathcal{O}(\omega^0)$ and $\mathcal{O}(\omega)$ are identical in the small-$\omega$ limit. As a result, at $\mathcal{O}(\omega)$, $\Ef_{rA}=0$ reduces to simply $\omega \Honef=0$, which gives $\Honef = \mathcal{O} (\omega)$. Similarly, one infers  $\Vf, \Rf= \mathcal{O} (\omega)$ from $\Ef_{tr}=0$ and $\Ef_{tA}=0$.
Thus, as long as we are interested in studying the properties of the  linearized perturbations at order $\mathcal{O}(\omega^0)$, we can effectively take $\Vf = \Rf = \Honef = 0$,\footnote{In contrast, odd-parity (axial) perturbations can have time independent velocity perturbations 
e.g., if the fluid is irrotational~\cite{Landry:2015cva}, which also affects odd-parity tidal perturbations~\cite{Pani:2018inf}.} which  will significantly simplify the form of the sources at second perturbative order. As a result,  the components $\Ef_{tr} $ and $\Ef_{tA}$ vanish identically at order $\mathcal{O}(\omega^0)$.

In the static limit, the remaining equations simplify considerably. We can now use $\Ef_{rA} = 0$ to  solve for $\Kf'$,
\begin{equation}
\label{adiabkp}
    \Kf' = \Hzerf\frac{2  \left(\Mm+4 \pi  r^3 \bar p\right)}{r (r-2 \Mm)}+\Hzerf' \;.
\end{equation}
Using this in $\gamma^{AB}\Ef_{AB} = 0$ (after substituting the pressure perturbation from $\Ef_{rr}=0$) we can solve for $\Kf$ in terms of $\Hzerf$ and $\Hzerf'$ only, obtaining 
\begin{widetext}
\begin{equation}
    \Kf = \Hzerf'\frac{2\left(\Mm+4 \pi  r^3 \bar p\right)}{\ell(\ell+1)-2}+\Hzerf \frac{ 2 r \Mm \left[4-\ell(\ell+1)+8 \pi  r^2 (3 \bar p+ \bar \rho
   )\right]+r^2 \left\{\ell(\ell+1) - 2+8 \pi  r^2 \left[ \bar p \left(8 \pi  r^2 \bar p-1\right)-\bar \rho \right]\right\}-4 \Mm^2 }{r
   \left(\ell(\ell+1)-2\right) (r-2 \Mm)} \; .
\label{adiabk}
\end{equation}
\end{widetext}

A further simplification occurs in the static limit: since perturbations evolve infinitely slowly compared to the rate of heat exchange, the fluid remains in local thermodynamic equilibrium effectively implying that perturbations are specified by a single fluid variable, i.e.,
\begin{equation}
\frac{\delta p}{p'} = \frac{\delta \rho}{\rho'} \;.
\label{adiabatic}
\end{equation}
This relation can be verified explicitly at first order. Using Eqs.~\eqref{adiabkp} and~\eqref{adiabk} in $\Ef_{rr} = 0$, one finds
\begin{equation}
\deltaf p = \frac{1}{2}(\bar p + \bar \rho)\Hzerf \; .
\end{equation}
Then, solving $\Ef_{tt}=0$ for $\deltaf \rho$, one obtains
\begin{equation}
\deltaf \rho = \frac{1}{2} \frac{\bar\rho'}{\bar p'} 
(\bar p + \bar \rho)\Hzerf \; .
\end{equation}
It is therefore natural to introduce the squared adiabatic sound speed,
\begin{equation}
c_s^2(r)\equiv\frac{\bar p' (r)}{\bar\rho'(r)}\;,
\end{equation}
so that $\deltaf p =  c_s^2 \deltaf \rho$.

With this definition in hand, substituting Eq.~\eqref{adiabk} into Eq.~\eqref{adiabkp} we obtain a single second-order differential equation for $\Hzerf$~\cite{Hinderer:2007mb},
\begin{widetext}
\begin{multline}
    \Hzerf''+\Hzerf'\frac{2 \left[r-\Mm+2 \pi  r^3 (\bar p- \bar \rho )\right]}{r (r-2 \Mm)}  \\
    +\Hzerf \left\{\frac{4 \pi  r (\bar p+ \bar \rho ) }{(r-2 \Mm)c_s^2}-\frac{2 \Mm r \left[- \ell(\ell+1)+52 \pi  r^2 \bar p+20 \pi  r^2 \bar \rho
   \right]+r^2 \ell(\ell+1)+4 \Mm^2+4 \pi  r^4 \left[ \bar p \left(16 \pi  r^2 \bar p-9\right)-5 \bar \rho \right]}{r^2 (r-2
   \Mm)^2}\right\}=0 \;.
   \label{master1}
\end{multline}
\end{widetext}

\subsection{Quadratic static equations}

To solve for the second-order variables, we expand Einstein equations to second order. These now take the form $\Es_{\mu \nu} = S_{\mu \nu}$. The left-hand side 
includes terms linear in the second-order perturbations (e.g., $\Hzers$, $\Hones$, etc.), 
 and is in form identical to $\Ef_{\mu \nu}$, with second-order variables replacing the linear ones. The right-hand side is a source term  accounting for the contributions quadratic in the linear variables. As at linear order, this tensor equation can be decomposed into components: $\Es_{tt} = S_{tt}$, $\Es_{tr} = S_{tr}$, and so on.

Expanding all perturbations in spherical harmonics and projecting the equations onto a particular multipole, the source terms involve integrals over products of three spherical harmonics and their derivatives. These integrals encode $SO(3)$ selection rules and are handled using straightforward integrations by parts, as described in App.~\ref{angIntApp} (see also~\cite{Iteanu:2024dvx,Pani:2013pma}).
Specifically, let ${}^{(1)}\!X_1^{\ell_1 m_1}$ and ${}^{(1)}\!X_2^{\ell_2 m_2}$ denote any of the linear perturbations (obtained by solving Eqs.~\eqref{Ett}-\eqref{EAB}) with harmonic numbers $\ell_1, m_1$ and $\ell_2, m_2$, respectively. The sources then take in general  the form
\begin{equation}
    S^{\ell m} (r) = \sum_{\ell_1, \ell_2, m_1,m_2} I^{m m_1 m_2}_{\ell \ell_1 \ell_2} S_{\ell \ell_1 \ell_2}
    \left[{}^{(1)}\!X_1^{\ell_1 m_1}(r), {}^{(1)}\!X_2^{\ell_2 m_2}(r) \right] \;,
\end{equation}
The sum is restricted to satisfy $m = m_1 + m_2$, $|\ell_1 -\ell_2|\le \ell \le |\ell_1+\ell_2|$, and $\ell\geq |m|$, with the integral vanishing unless 
 $\ell_1 + \ell_2 + \ell$ is even.
Additionally, the frequency of each second-order variable is in general given by the sum of the frequencies of the contributing linear modes.

We can now focus on the static limit. At linear order, we previously observed that $\Honef$, $\Vf$, and $\Rf$ vanish in this limit, and that $\Ef_{tr} = 0$ and $\Ef_{tA} = 0$ identically. We verified that, upon inserting the linear solutions in the zero-frequency limit, the corresponding second-order source terms satisfy $S_{tr} = 0$ and $S_{tA} = 0$. This implies that $\Hones$, $\Vs$, and $\Rsec$ also vanish in the static limit and therefore start at linear order in frequency, as for linear perturbations.\footnote{We note that, by explicitly solving for the second-order pressure and density perturbations in the static limit---following manipulations analogous to those used at linear order---one can verify that the condition~\eqref{adiabatic} continues to hold at second order. In particular, one finds
\be
\deltas p ( x^i) = c_s^2 \, \deltas \rho ( x^i) + \frac12 \frac{(c_s^2)'}{\rho'} \,  [ \deltaf \rho ( x^i) ]^2\;.
\ee
}

The procedure for deriving the second-order equation for $\Hzers$ parallels that used to obtain Eq.~\eqref{master1}, with the addition of the relevant source terms~\cite{Riva:2023rcm,Iteanu:2024dvx}. We find:
\begin{widetext}
\begin{multline}
    \Hzers''+\Hzers'\frac{2 \left[r-\Mm+2 \pi  r^3 (\bar p- \bar \rho )\right]}{r (r-2 \Mm)}  \\
    +\Hzers \Big\{\frac{4 \pi  r (\bar p+ \bar \rho ) }{(r-2 \Mm) c_s^2}-\frac{2 \Mm \left(-r \ell(\ell+1)+52 \pi  r^3 \bar p+20 \pi  r^3 \bar \rho
   \right)+r^2 \ell(\ell+1)+4 \Mm^2+4 \pi  r^4 \left[ \bar p \left(16 \pi  r^2 \bar p-9\right)-5 \bar \rho \right]}{r^2 (r-2
   \Mm)^2}\Big\}=S_{H_0} \;.
   \label{master2}
\end{multline}
\end{widetext}
To simplify the source $S_{H_0}$, we follow the procedure of~\cite{Riva:2023rcm,Iteanu:2024dvx}. Specifically, we use  angular integration identities, and the background and linearized Einstein equations. This allows us to express the source entirely in terms of the linear variable $\Hzerf(r)$, its radial derivative $\Hzerf'(r)$, and background quantities. We have verified that in the vacuum limit, i.e.~for $\bar \rho \to 0$, $\bar p \to 0$ and $\Mm(r) \to \text{const.}$, we recover the source derived in~\cite{Riva:2023rcm,Iteanu:2024dvx}.

For definiteness, we choose to perform the matching with the EFT for $\ell_1 = \ell_2 = \ell = 2$ and $m_1 = m_2 = m = 0$   \cite{Riva:2023rcm}.\footnote{Note that the quadratic coupling $\lambda_{222}$, like the linear Love number $\lambda_2$, does not depend on the magnetic quantum numbers \cite{Riva:2023rcm}. This is true more in general: one can easily see that there is only one independent coupling $\lambda_{\ell\ell_1\ell_2}$ for fixed multiplet $(\ell\ell_1\ell_2)$ at cubic order in perturbation theory~\cite{Bern:2020uwk,Iteanu:2024dvx,Combaluzier-Szteinsznaider:2024sgb}.}
In this case, the source reduces to
\begin{widetext}
\begin{align}
   - \left(I_{222}^{000} \right)^{-1} S_{H_0} =  & \  (\Hzerf')^2 + \Hzerf  \Hzerf' \frac{3   \left(\Mm
   +4 \pi  r^3 \bar  p\right)}{r(r-2
   \Mm)} + (\Hzerf)^2 \Big\{ \frac{\pi  r (\bar p+ \bar \rho )}{c_s^4(r-2\Mm)} \left( 1 + (c_s^2)'\frac{r(r-2\Mm)}{\Mm + 4\pi r^3 \bar p} \right) \nonumber \\ 
   &  +\frac{6 \pi  r  (\bar  p+ \bar  \rho ) }{(r-2
   \Mm) c_s^2}+\frac{  r \Mm \left[ 2 \pi  r^2 (71 \bar  p+15 \bar  \rho )+3\right]-7
   \Mm^2+\pi  r^4 \left[  \bar p \left(176 \pi  r^2 \bar  p-27\right)-15 \bar  \rho \right] +3 r^2 }{r^2 (r-2
   \Mm)^2} \Big\}
      \;, \label{source}
\end{align}
\end{widetext}
where $I^{000}_{222} = \frac{1}{7}\sqrt{\frac{5}{\pi}}$.



\section{Quadratic Love numbers of a neutron star}
\label{QLN}
In this section we present our results for the quadratic Love numbers of a neutron star. Although the analytical derivation previously presented is valid for the generic coupling between two external (electric) tidal fields with different angular momenta $\ell_1$ and $\ell_2$, here we focus on the most phenomenologically relevant case, $\ell_1=\ell_2=2$, and compute the quadratic correction to the quadrupolar ($\ell=2$) tidal response of the star, i.e.~$p_2$ (see Eq.~\eqref{dimscale}). The analysis for different choices of multipoles can be worked out in a similar way.

This problem was recently studied in Ref.~\cite{Pitre:2025qdf}, where $p_2$ was computed for neutron stars described by a polytropic EoS. Applying our independent approach, we reproduce their result and extend the analysis to realistic, nuclear-physics motivated EoS, for which the relation $p = p (\rho)$ is specified in tabulated form. This allows us to assess the behavior of  quadratic Love numbers for neutron-star models of direct astrophysical relevance.

\subsection{Numerical integration}
\label{sec:numint}
We numerically integrate the TOV equations for the background quantities $\left(\bar p(r),\bar\rho(r),\Mm(r),\Phi(r) \right)$, and then both Eqs.~\eqref{master1} and~\eqref{master2}, with the source of the latter given by Eq.~\eqref{source}. One can first solve the TOV equations, use the solution to integrate the first-order equation, then plug the solution into the source~\eqref{source} and, finally, numerically integrate Eq.~\eqref{master2}. However, in practice we numerically solve the entire system of equations simultaneously: this improves the numerical accuracy and avoids unnecessary interpolations of the variable. 

We start the integration near the origin, where we specify the central density $\rho_c \equiv \rho(r=0)$. Regularity implies $\Mm\simeq\frac{4\pi}{3}\rho_c r^3$, $\Phi\simeq{\rm const}$ and $p=p(\rho_c)$, the latter being fixed by the EoS. For the perturbation variables, regularity also requires $\Hzerf\simeq {\cal C}_1 \cdot r^2$ and $\Hzers\simeq {\cal C}_2 \cdot  r^2$, where ${\cal C}_1$ and ${\cal C}_2$ are arbitrary constants. As we will explain below, these constants can be chosen so that the exterior solution reproduces the correct amplitude of the tidal field at infinity. We then integrate numerically out to the stellar radius $\Rs$, defined by $p(\Rs)=0$,  and match the solution to the exterior.

The solutions form a one-parameter family, identified by the central density $\rho_c$ or, equivalently, by the compactness $C$, defined as
\be
C\equiv \frac{G m}{\Rs} \;.
\label{eq:defC}
\ee

\subsection{Matching at infinity and at the stellar surface}

We connect the interior solution found numerically to the exterior solution by imposing continuity\footnote{The matching of interior and exterior solutions is generally governed by Israel's junction conditions~\cite{Israel:1966rt}, which depend on how the pressure and energy density behave at the star's surface. For the EoS considered here---where pressure and energy density vanish smoothly at the surface---the junction conditions reduce to the continuity of the metric and its first derivative across the stellar boundary.  For related discussions on the implementation of junction conditions in the calculation of Love numbers, see also, e.g.,~\cite{Hinderer_2010, Pitre:2025qdf}.} of $H_0$ and its first radial derivative. To simplify notation, we will drop the subscript 0 from $H_0$, i.e.~$H \equiv H_0$.  First, we  consider linear perturbations and obtain the linear dimensionless Love number $k_2$; we then proceed to tackle the second order and obtain $p_2$.

\subsubsection{Linear perturbations}

The most general exterior solution of the linear equation~\eqref{master1} for $\ell =2$ is 
\begin{equation}
    \Hf_{\rm ext}(r) = \tilde{\mathcal{E}} [ H_{\uparrow}(r) +   \blin \, H_{\downarrow}(r) ] \;,
\end{equation}
where $\tilde  {\mathcal{E}}$ and $\blin$ are arbitrary constants, to be fixed by the boundary conditions, and we have defined the growing and decaying solutions as
\begin{align}
    H_{\uparrow}(r) & = r(r-r_s) \;, \\ H_{\downarrow} (r) & = \frac{5}{r_s^3}Q^{2}_2(2r/r_s-1) =   \frac{1}{r^3}(1+ \mathcal{O}(r_s/r)) \;,
\end{align}
respectively, with $Q^{2}_2$ being an associated Legendre function of the second kind. 
With this in hand, the $tt$-component of the linear metric perturbation outside the star is
\be
\deltaf g_{tt}^{(\ell=2,m)} (r) =  \tilde  {\mathcal{E}} \left(1-\frac{r_s}{r} \right) [ H_{\uparrow}(r) +   \blin \, H_{\downarrow}(r) ] \;.
\ee
This solution must be matched, for $r \gg \Rs  \ge r_s$, to the EFT linear solution, i.e.~to the first term on the right-hand side of Eq.~\eqref{eq:deltagttEFT}, yielding
\be
\label{matchinglin}
{\mathcal{E}}_{2,m} = \tilde  {\mathcal{E}} \;, \qquad \lambda_2 =  \frac{\blin}{12 G} \;.
\ee

We can now relate $\tilde {\mathcal{E}}$ and $\blin$ to the  numerical interior solution by matching at the stellar surface. By imposing continuity of $\Hf$ and its radial derivative at the surface, we obtain the following system of equations:
\begin{align}
    \Hf_{\rm int}(\Rs) &=  \tilde  {\mathcal{E}} [ H_{\uparrow}(\Rs) + \blin \, H_{\downarrow}(\Rs) ],\\
    \Hf'_{\rm int}(\Rs) &=  \tilde  {\mathcal{E}} [H'_{\uparrow}(\Rs) + \blin \, H'_{\downarrow}(\Rs)] \; .
\end{align}
We can then solve this system for $\tilde {\mathcal{E}}$ and $\blin$ in terms of $H_\uparrow(\Rs)$, $H_\downarrow(\Rs)$,  $\Hf_{\rm int}(\Rs)$ and $\Hf_{\rm int}'(\Rs)$, the latter two being determined by the numerical integration.
The value of $\tilde {\cal E}$ is proportional to the constant ${\cal C}_1$ of the regular interior solution at the center and fixes the tidal-field amplitude ${\cal E}_{2,m}$ as in Eq.~\eqref{matchinglin}; since it does not influence the matching, it can be chosen arbitrarily.
Furthermore, using the solution for $\blin$ together with the matching condition~\eqref{matchinglin} and the definition of $k_2$, Eq.~\eqref{dimscale},
yields~\cite{Hinderer:2007mb}
\begin{widetext}
\begin{equation}
    k_2 = \frac{8 C^5 (1-2 C)^2  [1+(1-2 C)(1 - y)]}{10 C \left[4 C^4-4 C^3+26
   C^2-24 C+\left(4 C^4+6 C^3-22 C^2+15 C-3\right) y+6\right]+15 (1-2
   C)^2 \log (1-2 C) [1+(1-2 C)(1 - y)]} \;,  \label{lambda2}
\end{equation}
\end{widetext}
where we have introduced the  logarithmic derivative at the stellar surface, 
\be
y \equiv   \frac{ \Rs\Hf'_{\rm int}(\Rs)}{\Hf_{\rm int}(\Rs)} 
\;. 
\label{eq:ydef}
\ee

For both the polytropic and realistic EoS considered here, $y$ varies slowly as the compactness changes across a reasonable range. Despite the $C^5$ factor in the numerator, Taylor-expanding $k_2$ for small compactness, while keeping $y$ fixed, yields
\begin{equation}
    k_2 =  -\frac{y-2}{2 (y+3)}+\frac{5 C (y (y+2)-6)}{2 (y+3)^2}+\mathcal{O}\left(C^2\right),
\end{equation}
showing that $k_2 = \mathcal{O}(1)$ in the $C\rightarrow0$ limit. For the realistic EoS considered here, one finds that $y \rightarrow 2$ in the small compactness limit, resulting in $k_2 \approx \mathcal{O}(1) C$.

Similarly, expanding $k_2$ around the black-hole compactness, $C = 1/2$, and keeping $y$ fixed, gives
\begin{equation}
    k_2 = \frac{4}{5} (C-1/2)^2+\mathcal{O}\left(\left(C-1/2\right)^3\right) \;,
\label{k2BH}
\end{equation}
which, interestingly, does not depend on $y$ at leading order. Although ordinary matter cannot support compact objects near the maximum compactness~\cite{Alho:2022bki}---thus
preventing from physically probing the black-hole limit---the formal expansion of  $k_2$ nevertheless captures the fact that $k_2 \rightarrow0$ as one approaches maximal compactness.

\subsubsection{Quadratic perturbations}
\label{matchingQuadratic}

We now proceed to match the second-order solution using the same approach as before, with the difference that, in addition to the homogeneous solutions, there is a contribution from the source term of the quadratic equation, Eq.~\eqref{master2}. To evaluate the source, Eq.~\eqref{source}, we use the linear solution computed above, for $(\ell=2 , m=0)$. 
The general second-order solution  outside the star can therefore be written as
\begin{equation}
    \Hs_{\rm ext}(r) =   {\mathcal{E}}_{2,0}^2 [ \cupa H_{\uparrow}(r) +  \cdown H_{\downarrow}(r) + H_{\rm part}(r) ],
\end{equation}
where $\cupa$ and $\cdown$ are integration constants and 
$H_{\rm part}(r)$ is the particular solution with the overall factor $\tilde  {\mathcal{E}}^2$ factored out. Without loss of generality, we choose the particular solution so that its large-$r$ expansion contains neither $r^2$ nor $1/r^3$ terms.  Focusing on the $(\ell=2 , m=0)$ multipole, this can be written as
\be
H_{\rm part}(r) = I_{222}^{000}  \left[ F_0(r) + G \lambda_2 F_1(r) + G^2 \lambda_2^2 F_2(r) \right] \;, 
\ee
where  
\begin{align}
F_{0} (r) & = - \frac12 r^4 \left( 1 + \frac12 \frac{r_s}{r} - \frac{3}{2} \frac{r_s^3}{r^3} \right) \;, \\
F_{1} (r) & = - \frac{12}{r}
\left[1- \frac{11 }{8 } \frac{ r_s}{r} +\mathcal{O}\left( \frac{r_s^3}{r^3}\right)  \right] \; , \\
F_{2} (r) & = - \frac{72}{r^6} \left[ 1
+\mathcal{O}\left( \frac{r_s}{r} \right)   \right]  \;.  
\end{align}
The explicit expressions of $F_{1,2} (r)$ are needed for the matching at the stellar surface and are given in App.~\ref{2ndOrderApp}, Eqs.~\eqref{partsol1} and ~\eqref{partsol2}.

The second-order $tt$ component of the metric,
\be
\deltas g_{tt}^{(\ell=2,m)} =   {\mathcal{E}}_{2,0}^2 \left(1 - \frac{r_s}{r} \right) [ \cupa H_{\uparrow}  +  \cdown H_{\downarrow}  + H_{\rm part}  ] \;,
\label{UVmetric}
\ee
must be matched, 
for $r \gg \Rs \ge r_s$, to the EFT quadratic solution, i.e.~to the second term on the right-hand side of Eq.~\eqref{eq:deltagttEFT}.
Choosing $\cupa$ such that the $r^2$ term  is absent from $\Hs_{\rm ext}$  guarantees that, at the leading order in $r_s/r$, the two metrics coincide.\footnote{Here we are assuming that ${\cal E}_{2,m} = \tilde {\cal E}$ also at second order. In principle, ${\cal E}_{2,m}$ and $ \tilde {\cal E}$ could differ at this order,  which would modify the condition on $\cupa$ without affecting the first-order matching conditions. A  discussion on the choice of the tidal field amplitude at second order can be found in~\cite{Pitre:2025qdf}.} 
Matching also the $1/r^3$ term, we obtain
\be
\lambda_{222} = -\frac{2}{63 G} \frac{\cdown}{I_{222}^{000}} \;.
\label{lambda222match}
\ee

To compute $\cdown$ we must impose the continuity of the metric and of its first derivative at the stellar boundary. The most general second-order solution for $H(r)$ inside the star  can be written as $\Hs_{\rm int}(r) + b \Hf_{\rm int}(r)$ where $\Hf_{\rm int}(r)$ and $ \Hs_{\rm int}(r)$ are, respectively,  solutions of Eqs.~\eqref{master1} and~\eqref{master2} in the star's interior, regular at the origin.\footnote{As discussed in Sec.~\ref{sec:numint}, when solving for $\Hf_{\rm int}(r)$ and $\Hs_{\rm int}(r)$ numerically we must specify values for ${\cal C}_1$ and ${\cal C}_2$, which correspond to particular choices of $\mathscr{E}_{2,0}$ and $b$. However, neither $y$, $z$ nor $p_2$ depend on this choice.} These two solutions are obtained numerically, as detailed in Sec.~\ref{sec:numint}. Matching the interior and exterior solutions at the boundary gives 
\begin{align}
  &  \Hs_{\rm int}(\Rs) + b\, \Hf_{\rm int}(\Rs) = \nonumber \\ & \qquad \mathcal{E}_{2,0}^2 [ \cupa H_{\uparrow}(\Rs) + \cdown H_{\downarrow}(\Rs) + H_{\rm part}(\Rs)] \;, \\
  &  \Hs'_{\rm int}(\Rs) +b\, \Hf'_{\rm int}(\Rs) = \nonumber \\  & \qquad \mathcal{E}_{2,0}^2[ \cupa H'_{\uparrow}(\Rs) + \cdown H'_{\downarrow}(\Rs)+H_{\rm part}'(\Rs)] \;,
\end{align}
yielding a system of equations for $b$ and $\cdown$. Solving this system and using Eqs.~\eqref{lambda222match}  and~\eqref{dimscale} allows us to obtain an expression for $p_2$ in terms of the compactness $C$, of $y$ defined in Eq.~\eqref{eq:ydef}, and the dimensionless variable $z$, defined as
\begin{equation}
z \equiv \frac{ r_s}{2 I^{000}_{222} \, \Hf_{\rm int}(\Rs)}\left( \frac{\Hs_{\rm int}(\Rs)}{\Hf_{\rm int}(\Rs)} \right)' \;.
\end{equation}
The full expression for 
$p_2$ is rather lengthy and is given in App.~\ref{2ndOrderApp}. Its expansion around $C=0$, while keeping $y$ and $z$ fixed, is given by
\begin{equation}
    p_2 = -\frac{50 z}{7 (y+3)^3} - \frac{25 C [ y (y+3)-2 z(5 y+24)]}{7 (y+3)^4}+ \mathcal{O}(C^2).
\end{equation}
For the realistic EoS considered here, $z$ changes approximately linearly across most of the relevant compactness range, and approaches to a constant value in the limits of zero and maximum compactness. As in the linear-order case, a nontrivial cancellation occurs between powers of $C$ in the numerator and denominator. For realistic EoS, one finds $y \to 2$ and $z \to 0$ in the small compactness limit, resulting in $p_2 \rightarrow 0$.
(Numerical results are shown in Fig.~\ref{fig:p2-222} in the next section.)

Formally expanding around the black-hole limit yields
\begin{equation}
    p_2 = \mathcal{O}\left(\left(C-1/2\right)^2\right) \;,
    \label{eq:BHlimit}
\end{equation}
which---as discussed---is not physically accessible for ordinary neutron stars, but correctly captures the fact that $p_2 \to 0$ in the black-hole limit~\cite{Riva:2023rcm}. To compute the exact numerical factor in Eq.~\eqref{eq:BHlimit} one would need to include the $(r_s/r)^2$-corrections neglected in the matching  discussed at the end of Sec.~\ref{sec:EFTquadrupole}. However, since these corrections scale as $C^3 k_2(C)$, Eq.~\eqref{k2BH} shows that they cannot alter the overall scaling $p_2 \sim (C-1/2)^2 $ in the large compactness limit.

To summarize, in order to compute the linear and quadratic Love numbers, one needs the logarithmic derivative $y$ and $z$, at first and second order, respectively. These quantities (together with the mass $m$ and radius $R_*$ of the configuration) allow computing $\lambda_2$ and $\lambda_{222}$---or, equivalently, the corresponding dimensionless quantities, $k_2$ and $p_2$, see Eq.~\eqref{dimscale}---via Eq.~\eqref{lambda2} and Eq.~\eqref{lambda222}, respectively.

\subsection{Results}
\label{sec:numres}

\begin{figure*}[t]
\begin{center}
\includegraphics[width=0.4275\textwidth]{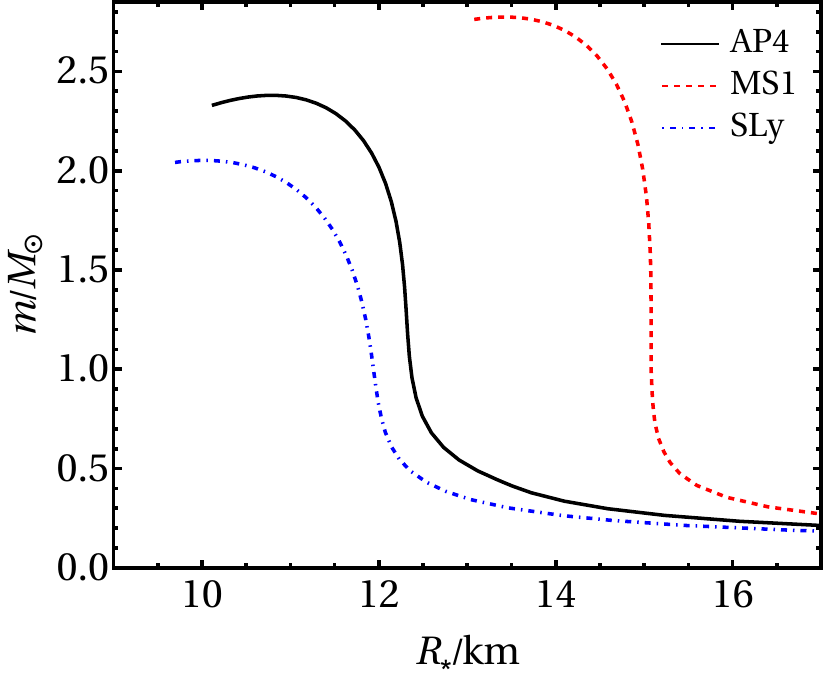}
\hspace{1cm}
\includegraphics[width=0.437\textwidth]{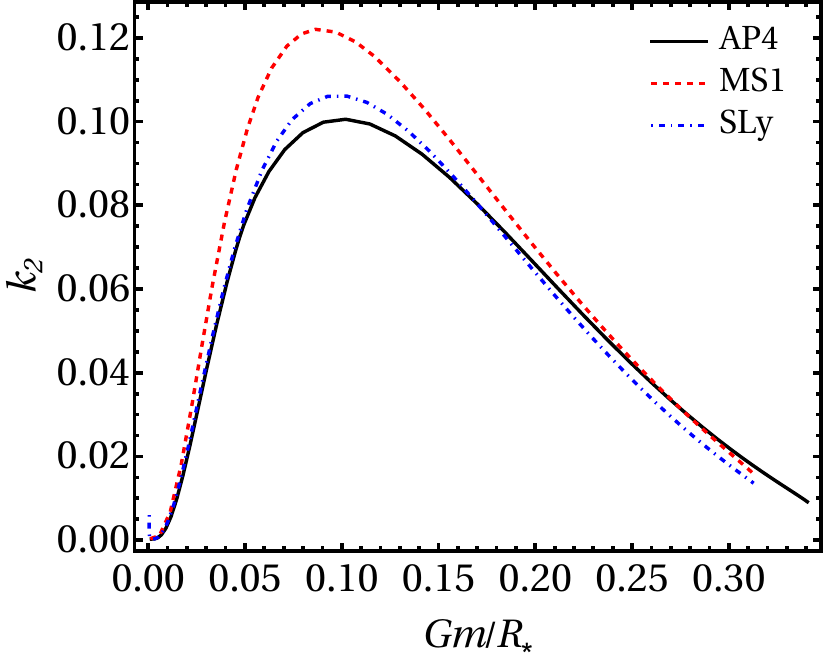}
\end{center}
\caption{
Left: Mass-radius diagram for the neutron-star EoS considered in this work.
Right: The corresponding linear tidal Love number as a function of compactness $C= Gm/\Rs$.}
\label{fig:MR}
\end{figure*}
\begin{figure*}[t]
\begin{center}
    \includegraphics[width=0.437\textwidth]{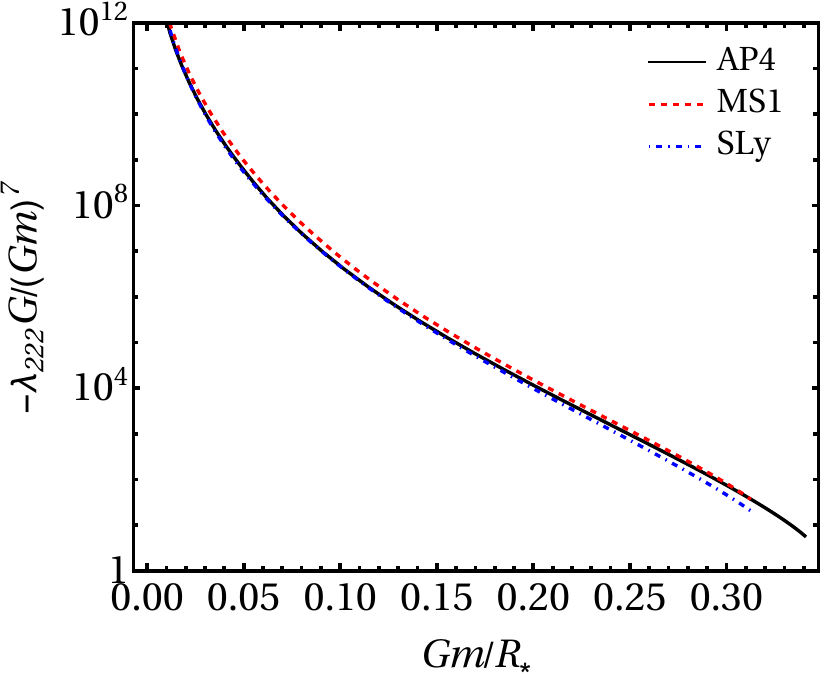}
    \hspace{1cm}
	\includegraphics[width=0.4275\textwidth]{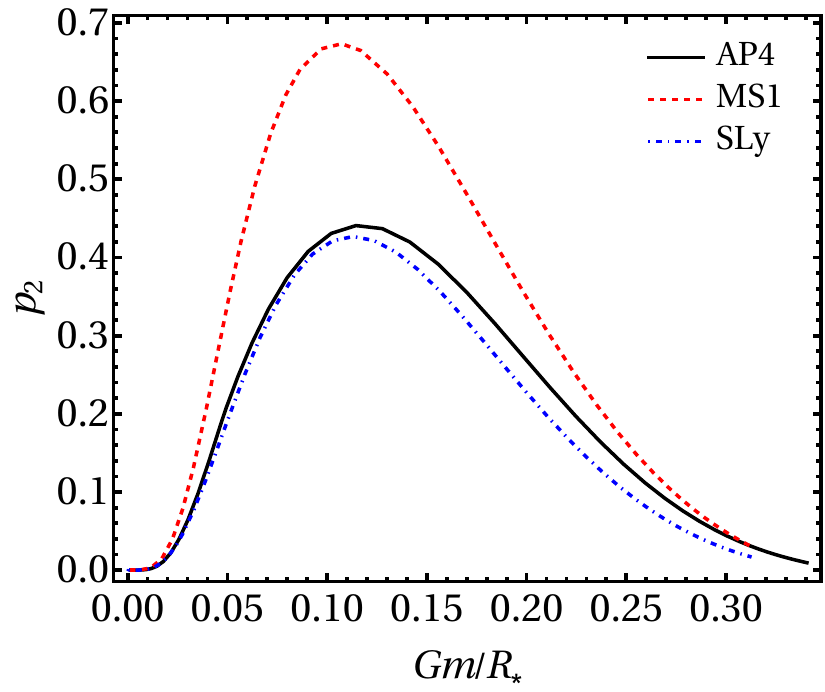}
\end{center}
\caption{Left: Quadratic Love number normalized as $G \lambda_{222} (Gm)^{-7}$ as a function of the compactness,    for different EoS.
Right: Dimensionless coefficient $p_2$ capturing the leading quadratic tidal deformation as a function of compactness,  for different EoS.}
\label{fig:p2-222}
\end{figure*}
To validate our procedure and code, we have preliminary checked that we  reproduce the results of~\cite{Pitre:2025qdf} for a polytropic EoS of the form
\begin{equation}
    \rho(p)=\left(\frac{p}{K}\right)^{n/(n + 1)} + n p\,,
\end{equation}
where $K>0$ and $n>0$ are constant.
In addition, we have considered tabulated, nuclear-physics-based EoS that cover a wide range of neutron-star deformabilities and allow for neutron stars with maximum mass larger than $2M_\odot$, as required by pulsar-timing observations (e.g.,~\cite{Antoniadis:2013pzd,NANOGrav:2019jur}).
In particular, we consider APR~\cite{Akmal:1998cf}, 
MS1~\cite{Mueller:1996pm},
SLy4~\cite{Douchin:2001sv}. The mass-radius diagram and the usual linear tidal deformability of these EoS are given in Fig.~\ref{fig:MR}.

The values for the quadratic Love number $\lambda_{222}$ and the corresponding dimensionless coefficient $p_2$---defined in Eq.~\eqref{dimscale}---for these EoS are shown in Fig.~\ref{fig:p2-222}. In the relevant regime of compactness, we observe that $\lambda_{222}$ (suitably normalized by powers of $G$ and $m$) is ranging from $\mathcal{O}(10^3)$ to $\mathcal{O}(10^7)$. For the EoS we consider, we find $p_2={\cal O}(0.1)$ in the relevant regime. Furthermore, we observe that different EoS predict qualitatively the same dependence of $p_2$ on compactness. In the limit of maximal compactness for a given EoS, $p_2$ approaches the black-hole value $p_2^{\rm BH}
= 0$~\cite{Riva:2023rcm,Iteanu:2024dvx}.
The cancellation in the full expressions for $\lambda_{222}$ and $p_2$ discussed in Sec.~\ref{matchingQuadratic} makes it difficult to evaluate them numerically in the small compactness limit. In that regime, we use the Taylor expansion of the full result around  $C = 0$. For small compactness, we find $p_2 \rightarrow0$, in contrast with polytropic EoS where $p_2$ approaches a constant value~\cite{Pitre:2025qdf} (note, however, that tabulated EoS are not meant to describe the small compactness regime and should be matched to an EoS that is valid in that limit). Nevertheless, the dimensionful quadratic Love number $\lambda_{222}$ grows in the small compactness limit due to the $R_*^{8}$ factor relating it to  $p_2$.

\begin{figure}[t]
\begin{center}
	\includegraphics[width=0.42\textwidth]{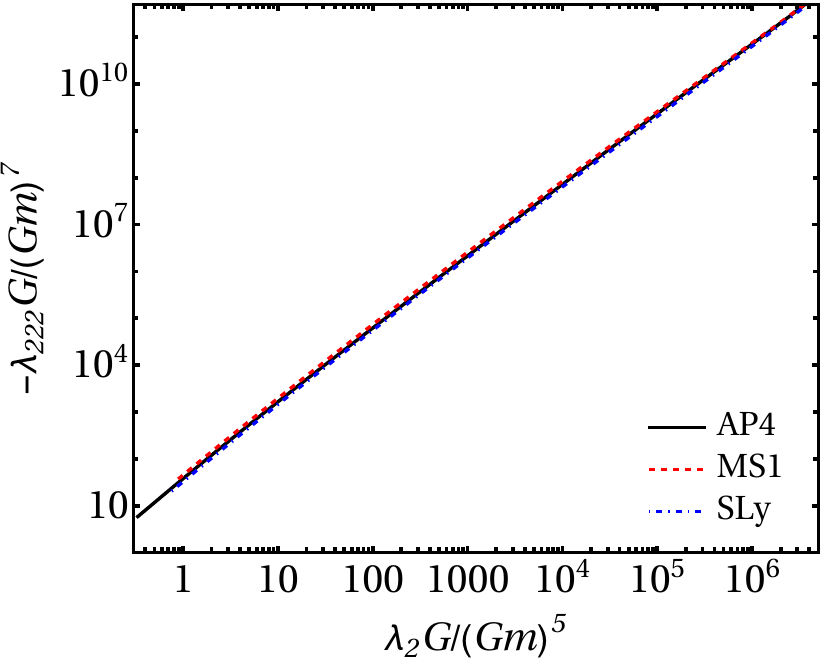}\\
    \includegraphics[width=0.425\textwidth]{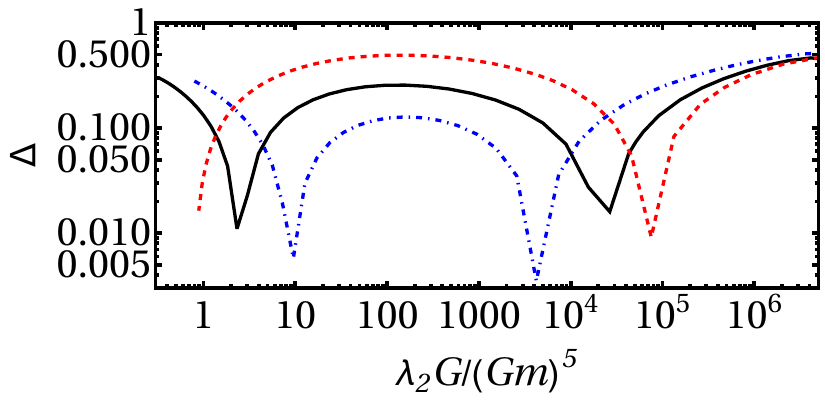}
\end{center}
\caption{
Top panel: The approximately EoS-independent relation among the quadratic and linear Love numbers, normalized by suitable powers of the mass.
Bottom panel: relative deviation from the fit~\eqref{universal}, i.e. $\Delta=\left|1-{Y}/{Y_{\rm fit}}\right|$.
}
\label{fig:universal}
\end{figure}

Given the quadratic Love numbers, it is interesting to investigate whether they feature some approximately universal relation that is only mildly sensitive to the EoS, in analogy with their linear counterpart~\cite{Yagi:2013sva,Yagi:2016bkt}.
In Fig.~\ref{fig:universal}, we consider the relation between linear and quadratic Love numbers, both normalized by suitable powers of $G$ and $m$. Interestingly, the relation is only mildly sensitive to the EoS. In a log-log scale, it is almost linear and well fitted by
\begin{equation}
    Y_{\rm fit} = 3.77 +1.51 X+0.005 X^2 \;,\label{universal}
\end{equation}
where $Y=\log \left(-G\lambda_{222}/(Gm)^7\right)$ and $X=\log \left(G\lambda_2/(Gm)^5\right)$. The bottom panel of Fig.~\ref{fig:universal} shows that this fit is accurate at least within $50\%$ (but typically more accurate than that) for the EoS considered here, which spans a large range in the mass-radius diagram (see Fig.~\ref{fig:MR}).

\section{Quadratic tidal effects in the waveform}
\label{sec:waveform}
\begin{figure}[t]
\begin{center}
	\includegraphics[width=0.48\textwidth]{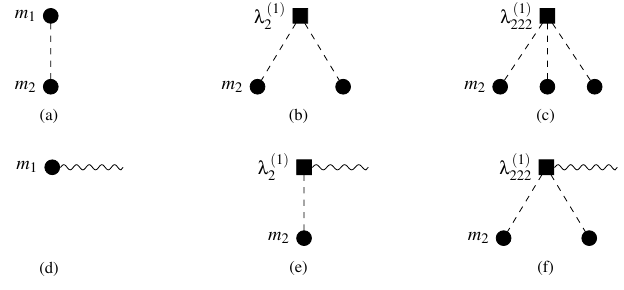}
\end{center}
\caption{Feynman diagrams needed to compute the contribution of the nonlinear Love number to the waveform. Dashed lines correspond to exchange of potential gravitons propagating instantaneously while wavy lines represent radiated graviton. Diagrams (a), (b) and (c) contribute to the bound energy, while (d), (e) and (f) are needed to compute the energy flux. One should add the equivalent diagrams given by the swapping of the two objects to obtain the full result.} 
\label{fig:WaveD}
\end{figure}

In this section, we compute the leading effect of the quadratic Love number $\lambda_{222}$ on the waveform. Following a standard procedure~\cite{Goldberger:2004jt,Blanchet:2013haa}, we restrict our attention to a binary system of compact objects in a circular orbit, with masses $m_1$ and $m_2$ and total mass 
\be
M \equiv m_1 + m_2 \;.
\ee

\subsection{Binding energy, fluxes, and GW phase}

We first compute the total contribution of tidal effects to the binding energy $E_{\rm B}$. We then compute the energy flux carried away by the GW, $\mathcal{F}_E$. For circular orbits, these two quantities are functions of the orbital frequency $\omega$. Introducing the standard dimensionless PN parameter $x \equiv (G M \omega)^{2/3}$, we can write the evolution of the orbital phase $\phi$ of the binary as
\begin{align}
	\frac{\D x}{\D t} = -\frac{\mathcal{F}_E(x)}{\D E_{\rm B}(x)/\D x}
	\; , 
	\qquad
	\frac{\D \phi}{\D t} = \frac{x^{3/2}}{G M } \; .
	\label{eq:xandt}
\end{align}

In order to compute $E_{\rm B} (x)$ and $\mathcal{F}_E(x)$, we use the EFT framework of Ref.~\cite{Goldberger:2004jt} (see, e.g.,~\cite{Porto:2016pyg,Riva:2019xjs,Goldberger:2022rqf} for reviews). In particular, we consider two objects described by the actions~\eqref{eq:Spp} and~\eqref{eq:Stidal} and split the gravitational field into a potential mode and a radiation mode. We then introduce the usual diagrammatic conventions for the propagators of the potential and radiation modes, i.e.,
\begin{align*}
	\raisebox{3pt}{\PotConv} & \equiv \ \text{potential graviton} \, , \\
	\raisebox{3pt}{\gravconv} & \equiv \ \text{radiation graviton} \, . 
\end{align*}
The binding energy can be obtained by integrating out the potential modes, i.e., by evaluating all tree-level diagrams built solely from potential-field propagators. For our purposes, we require only the leading Newtonian contribution. This is obtained from diagram (a) in Fig.~\ref{fig:WaveD}, together with the linear and nonlinear tidal-response contributions represented by diagrams (b) and (c) of the same figure, respectively.  Explicitly, we find
\begin{equation}
    E_{\rm B} = -\frac{G m_1 m_2}{\rorb} - \frac{6 G^2 \lambda^{(1)}_2 m_2^2}{\rorb^6} +\frac{6 G^3 \lambda^{(1)}_{222} m_2^3}{\rorb^9} + (1 \leftrightarrow 2) \; ,
    \label{eq:BindEn}
\end{equation}
where we have introduced the orbital distance $\rorb$. Here, $\lambda^{(1)}_2$ and $\lambda^{(1)}_{222}$ are respectively the linear and nonlinear Love numbers of the object $m_1$.

To determine the flux, we evaluate the pseudo stress-energy tensor $T^{\mu\nu}$ that incorporates the nonlinear gravitational contributions of the potential gravitons generated by the Einstein–Hilbert action and the worldline couplings. Its coupling to the radiation graviton $\bar h_{\mu \nu}$ is given  by 
\be
- \frac{1}{2 \Mpl} \int \dd^3 x \, T^{\mu \nu} \bar h_{\mu \nu} \;.
\ee
In particular, we extract the leading PN contributions to $T^{\mu \nu} $ arising from the point-particle action and from the linear and quadratic tidal deformations, corresponding to diagrams (d), (e), and (f) in Fig.~\ref{fig:WaveD}, respectively. With this at hand, we can then compute the total quadrupole moment of the binary $I^{ij}$ as follows
\begin{align}
    I^{ij} & = \int \dd^3 x \; x^{\langle i}x^{j\rangle} T^{00} \notag \\
    & = \left[\frac{m_1 m_2}{M} \rorb^2  + 12\lambda_2^{(1)} \frac{G m_2}{\rorb^3} - 18 \lambda_{222}^{(1)}\frac{G^2 m_2^2}{\rorb^6}  + (1 \leftrightarrow 2) \right] n_{\rm o}^{\langle i}n_{\rm o}^{j\rangle},
\end{align}
where angular brackets denote trace-subtracted symmetrization of the enclosed indices and $n^i_{\rm o}$ is the unit vector pointing from one object in the binary system to the other.
Given the quadrupole, we can compute the total energy flux using the usual quadrupole formula
\begin{equation}
    \mathcal{F}_E = \frac{G}{5} \dddot{I}_{ij} \dddot{I}^{ij} \; .
\end{equation}

For simplicity, we focus on quasi-circular orbits. 
Following Refs.~\cite{Henry:2020ski, Mandal:2024iug}, we  introduce the dimensionless symmetric and antisymmetric combinations of the linear Love number,\footnote{Because  Refs.~\cite{Henry:2020ski, Mandal:2024iug} use a different normalization convention for $\lambda_2$, $\lambda_2^{\rm there} = 4 \lambda_2^{\rm here}$, we multiply our definition of $\tilde \lambda_\pm$ by an additional factor of 
4 to match theirs. } 
\begin{equation}
	\tilde{\lambda}_{\pm}  \equiv \frac{2}{G^4 M^5}\left(
	\frac{m_2}{m_1}\lambda_2^{(1)} \pm \frac{m_1}{m_2}\lambda_2^{(2)}
	\right) \; .
    \label{lambdapm}
\end{equation}
Analogously, for the quadratic Love number we define
\be
\tilde{\rho}_{\pm}  \equiv \frac{2}{G^6 M^7}\left(
	\frac{m_2}{m_1}\lambda_{222}^{(1)} \pm \frac{m_1}{m_2}\lambda_{222}^{(2)}
	\right) \; .
    \label{rhopm}
\ee
Then, using the binding energy~\eqref{eq:BindEn}, we can find the radial separation of the binary as a function of the PN parameter $x$,
\begin{equation}
	\rorb = \frac{G M}{x}\left[
	1+6x^5 \tilde{\lambda}_+  -\frac{9}{2}x^8(\tilde{\rho}_+ -\delta \tilde{\rho}_-)
	\right] \; .
\end{equation}
Similarly, we  find the explicit expressions for the bound energy and the emitted flux,
\begin{align}
	E_{\rm B} & = -\frac{M \nu x}{2}\left[
	1 - 18x^5 \tilde{\lambda}_+  + 15 x^8 (\tilde{\rho}_+ -\delta \tilde{\rho}_-)
	\right] \, , \\
	\mathcal{F}_E & = \frac{32\nu^2 x^5}{5 G}
	+ \frac{192\nu x^{10}}{5 G}[
	\tilde{\lambda}_+(1+4\nu) + \delta \tilde{\lambda}_-
	] \notag \\
	& \quad -\frac{288 \nu^2 x^{13}}{5 G}(
	4\tilde{\rho}_+ - 2\delta \tilde{\rho}_-
	)  \; ,
\end{align}
where $\nu$ and $\delta$ are the usual dimensionless symmetric mass ratio and mass-difference parameters, respectively defined as
\begin{equation}
	\nu \equiv \frac{m_1 m_2}{M^2}\; ,
	\qquad
	\delta \equiv \frac{m_1 - m_2}{M} \; .
\end{equation}
Notice that the contribution from the linear tidal effects agrees with the leading-order result previously computed in~\cite{Flanagan:2007ix,Mandal:2023lgy, Mandal:2024iug}.

Using these results together with the flux–balance relation~\eqref{eq:xandt}, we find that the time-domain half-phase of the dominant quadrupole $(\ell,m) = (2 , \pm 2)$ mode is given by
\begin{equation}
    \phi(x) = \phi_0 -\frac{x^{-5/2}}{32 \nu} + \phi_{\rm LL}(x) + \phi_{\rm QL} (x) \;,
   \label{orbphase}
\end{equation}
where $\phi_0$ is an integration constant and
\begin{align}
	\phi_{\rm LL}(x)  & =   -\frac{3 x^{5/2}}{16 \nu ^2}\left[ \tilde{\lambda}_+ (1 + 22 \nu) + \delta  \tilde{\lambda}_-\right] \;, \\ 
    \phi_{\rm QL} (x) & = 
	\frac{45 x^{11/2}}{352 \nu}\left(19 \tilde{\rho}_+-17 \delta  \tilde{\rho}_-\right) \; .
\end{align}
It is straightforward to extend this computation by including all PN corrections in the point-particle approximation, in which case one obtains $\phi = \phi_0 +\phi_{\rm pp} + \phi_{\rm LL} + \phi_{\rm QL} $, where $\phi_{\rm pp}$ is the point-particle time-domain half-phase~\cite{Blanchet:2013haa}.

To connect with observations, we compute the waveform phase in the frequency domain, $\psi(f)$, as a function of the Fourier frequency
 $f$, using the stationary phase approximation. This is given by \cite{Cutler:1994ys}
\begin{equation}
  \psi(f) = 2 \pi f t(f) - 2\phi(f)  -\frac{\pi}{4}\;,
  \label{eq:Psi-Tidal}
\end{equation}
where $t(x)$ and $\phi(x)$ are determined from the energy-balance equations, Eq.~\eqref{eq:xandt}.
We obtain 
\begin{equation}
	\psi(f)   = 2 \pi f t_c + \psi_c - \frac{\pi}{4} + \psi_{\rm pp}(f) + \psi_{\rm LL}(f) + \psi_{\rm QL}(f) \; , \label{finalwaveform}
\end{equation}
where $t_c$ and $\psi_c$ are the coalescence time and phase, respectively, and $\psi_{\rm pp}$ is the phase in the point-particle approximation. The leading linear and quadratic tidal contributions to the phase are, respectively,
\begin{align}
	\psi_{\rm LL}(f)  & = -  \frac{9 \v^{5}}{16 \nu^2}\left[
	\tilde{\lambda}_+(1+22\nu) + \tilde{\lambda}_- \delta
	\right] \;, \\ 
    \psi_{\rm QL} (f) & = 
	\frac{135 \v^{11}}{1408 \nu}\left(
	19 \tilde{\rho}_+ -17 \tilde{\rho}_- \delta 
	\right) \; ,
\end{align}
with $\v  \equiv (\pi G M f)^{1/3}$. Again, $\psi_{\rm LL}$ agrees with what was previously computed in the literature~\cite{Flanagan:2007ix,Henry:2020ski}, while $\psi_{\rm QL}$ is one of our main results.

\begin{figure*}[t]
\begin{center}
	\includegraphics[width=0.33\textwidth]{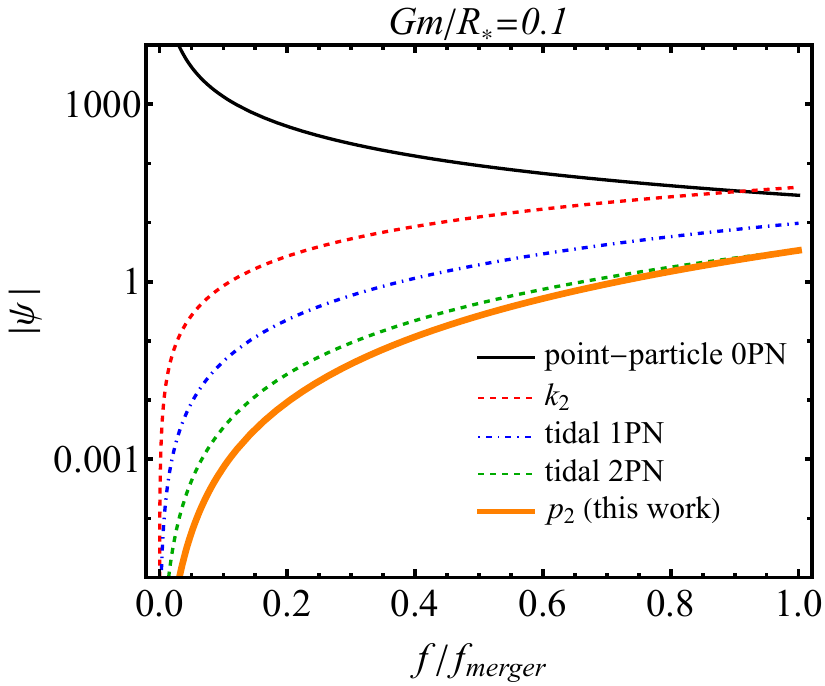}
    \includegraphics[width=0.33\textwidth]{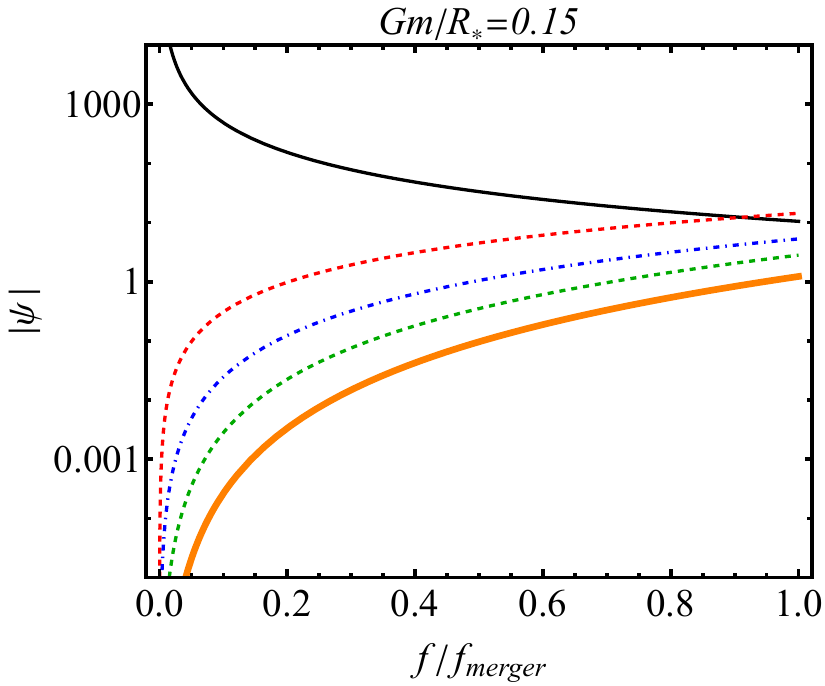}
    \includegraphics[width=0.33\textwidth]{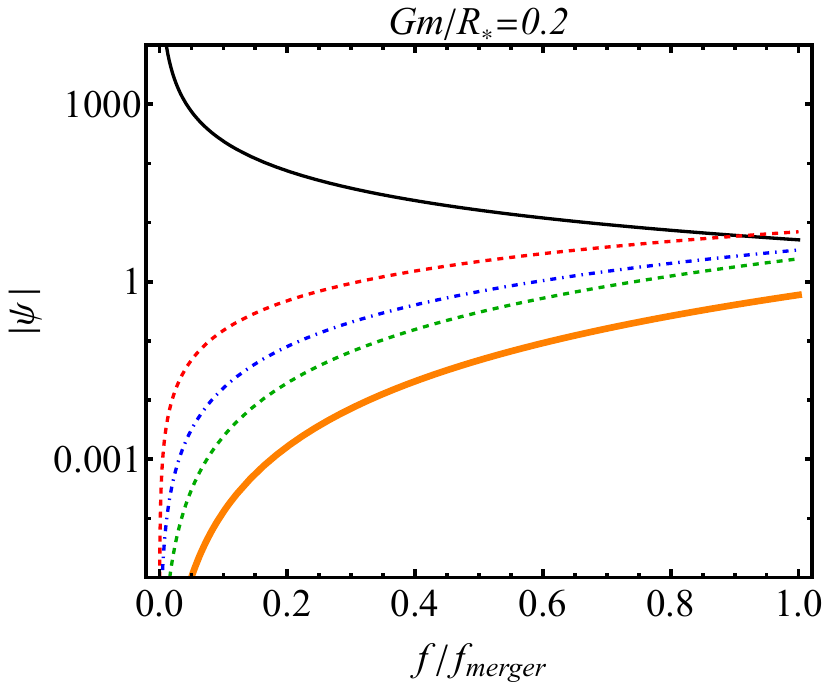}
\end{center}
\caption{Comparison between different contributions to the GW phase $\psi$ as function of the frequency, rescaled by the Newtonian estimate for the merger frequency (see Eq.~\eqref{fmerger}).
We consider a reference circular binary with $m_1=m_2=1.4M_\odot$, linear Love numbers $k_2=0.1$, and quadratic Love number $p_2=0.4$.
The black continuous curve is the leading-order point-particle phase, the red, blue, and green curves correspond to the leading, next-to-leading, and next-to-next-to-leading corrections due to the linear tidal Love number $k_2$. The orange thicker curve is the new contribution coming from the leading quadratic Love number computed in this work.
Each panel corresponds to a different compactness. As a reference, $C=(0.1,0.15,0.2)$ respectively corresponds to $k_{2}/C^5\approx(10\, 000,1317,313)$, $R_*\approx(20.7,13.8,10.4)\,{\rm km}$, and $f_{\rm merger} \approx (730, 1341, 2065)\,{\rm Hz}$.
}
\label{fig:phase}
\end{figure*}
\begin{figure*}[t]
\begin{center}
	\includegraphics[width=0.330\textwidth]{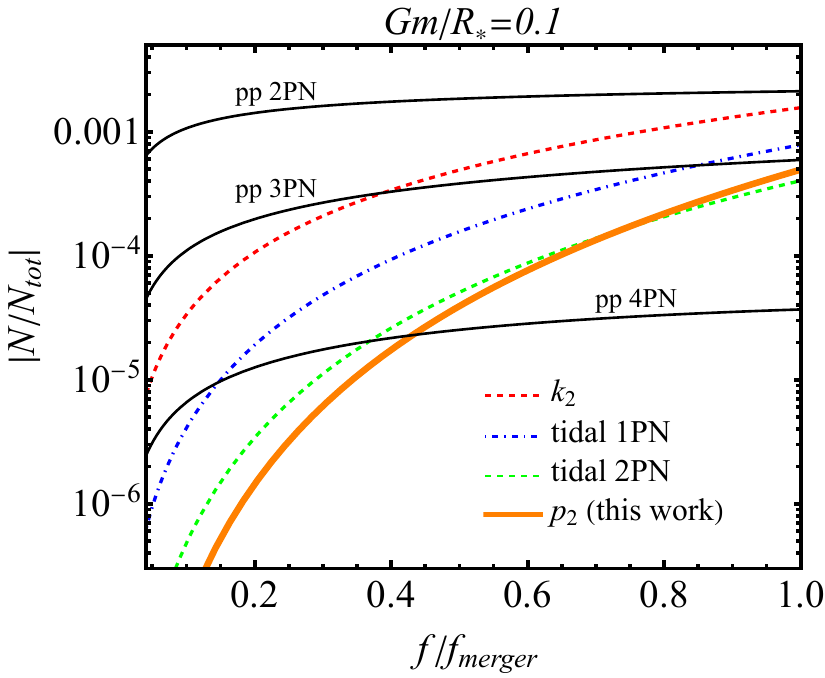}
    \includegraphics[width=0.330\textwidth]{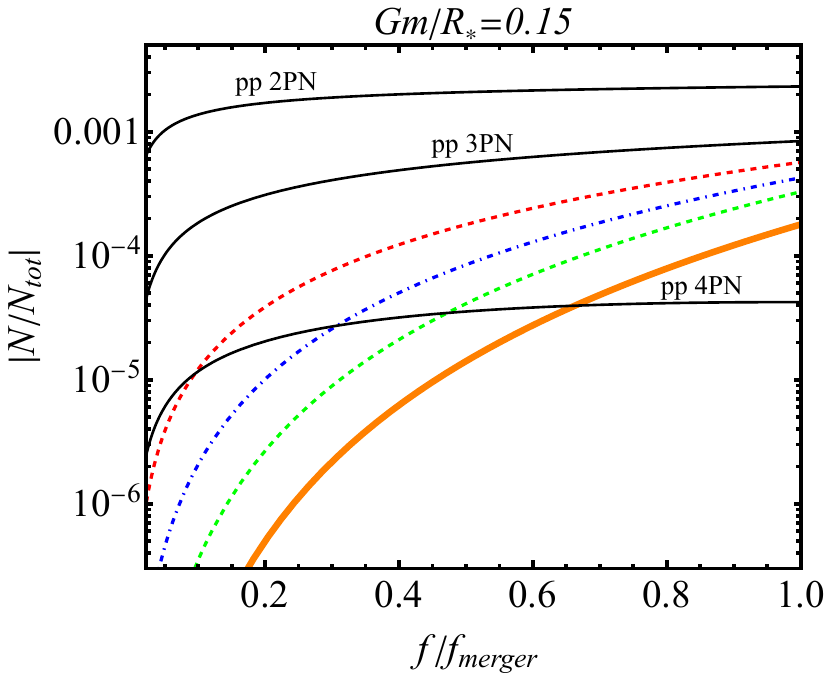}
    \includegraphics[width=0.330\textwidth]{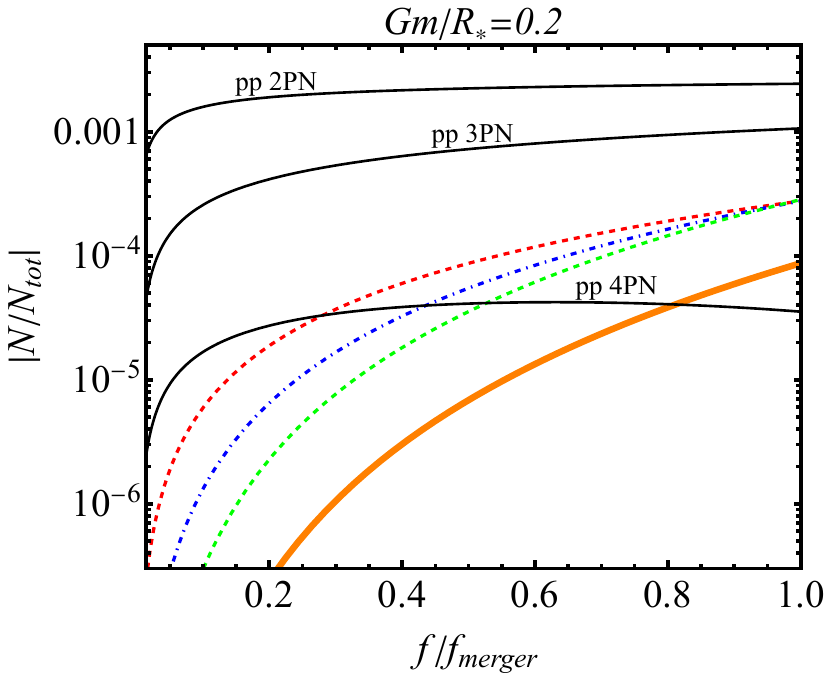}
\end{center}
\caption{Comparison between different contributions to the GW cycles $N( f) \equiv [\phi(f) - \phi(f_{\rm entry})]/\pi$, where here $f = \omega/\pi$  and $f_{\rm entry}=30\,{\rm Hz}$ for reference, as function of the frequency rescaled by the Newtonian estimate for the merger frequency (see Eq.~\eqref{fmerger}), normalized by the total number of cycles.
We considered the same binary parameters as in Fig.~\ref{fig:phase}.
The black continuous curves are the 2PN, 3PN and 4PN point-particle contributions, respectively, starting from above.
The red, blue, and green curves correspond to the leading, next-to-leading, and next-to-next-to-leading corrections due to the linear tidal Love number $k_2$. The orange thicker curve is the new contribution coming from the leading quadratic Love number computed in this work.
}
\label{fig:cycles}
\end{figure*}

\subsection{Estimates}
We can now estimate the relative impact of the quadratic Love number on the waveform.  
Since $\psi_{\rm pp} \propto \v^{-5}$  at leading-order in $\v$, Eq.~\eqref{finalwaveform} shows explicitly that the quadratic Love numbers enter the GW phase at 8PN order, i.e.~they are suppressed by a relative 3PN factor compared to the leading-order tidal term.  
However, in Sec.~\ref{sec:numres} we showed that $p_{2} \sim 0.5$ for a range of realistic EoS, and the coefficients actually entering the waveform, $\tilde \rho_\pm$, are enhanced by  negative powers of the compactness, $\tilde \rho_\pm \sim p_2/C^8$, see Eqs.~\eqref{dimscale} and~\eqref{rhopm}.

In Fig.~\ref{fig:phase} we compare the various contributions to the GW phase for a reference circular binary with $m_1 = m_2 = 1.4\,M_\odot$. In this case,  
Eq.~\eqref{finalwaveform} yields 
\begin{equation}
    \frac{\psi_{\rm QL}}{\psi_{\rm LL}}
   \approx 0.1 \cdot \frac{p_{2}/k_{2}}{5} \left( \frac{C}{0.1} \right)^{-3}\left(\frac{f}{730\, \text{Hz}} \right)^2\,.
    \label{backoftheenvelope}
\end{equation}
Since $p_{2}$ and $k_{2}$ are of comparable magnitude (actually, $p_2\approx 5 k_2$, see Figs.~\ref{fig:MR} and~\ref{fig:p2-222}), this estimate confirms that the relative importance of the quadratic Love number increases in the small-compactness limit, which may partially compensate for its nominal 3PN suppression.
The left, middle, and right panels correspond to compactness values $C = (0.1,\,0.15,\,0.2)$, respectively.

In Fig.~\ref{fig:cycles}, we compare the various contributions to the accumulated GW cycles, $N(f)=[\phi(f)-\phi(f_{\rm entry})]/\pi$,  with $f = \omega/\pi$  and where $f_{\rm entry}=30\,{\rm Hz}$ is taken as a reference initial frequency. Each contribution is normalized by the total number of cycles $N_{\rm tot}(f)$. We display only the even-PN contributions to both the point-particle approximation and the linear tidal effects. The total number of cycles used for the normalization, however, includes all contributions up to 4.5PN order in the point-particle approximation and 2PN order in the linear tidal effect.

Overall, the relative importance of tidal terms shown in Figs.~\ref{fig:phase} and~\ref{fig:cycles} 
depends significantly on both the compactness and the GW frequency, increasing towards the merger. 
At relatively small compactness ($C\approx0.1$), the linear tidal terms eventually become comparable to the point-particle phase near the merger.
In Figs.~\ref{fig:phase} and~\ref{fig:cycles} we normalize the horizontal axis by the Newtonian estimate for the merger frequency (i.e., at $\rorb = 2\Rs$) which, for an equal-mass binary, yields $f_{\rm merger} = \frac{\sqrt{C}}{2\pi \Rs}$. Assuming $m_1=m_2=1.4 M_\odot$, this gives
\begin{equation}
    f_{\rm merger} = 730 \, \text{Hz} \, \left(\frac{C}{0.1}\right)^{3/2} \,,\label{fmerger}
\end{equation}
corresponding to $f_{\rm merger} = (730, 1341, 2065)\,{\rm Hz}$ for $C=(0.1, 0.15, 0.2)$, respectively. Note that this is just a rough estimate since, when the compactness is large,
the PN parameter $x \approx C (f/f_{\rm merger})^{2/3}$ can be sizable already before the merger, signaling a breakdown of the PN expansion at higher frequencies.

Evaluating Eq.~\eqref{backoftheenvelope} using the expression above for the merger frequency, we find
\be
   \frac{\psi_{\rm QL}}{\psi_{\rm LL}} 
   \approx 0.1 \cdot \frac{p_{2}/k_{2}}{5} \left( \frac{f}{f_{\rm merger}} \right)^2 \,.
\ee
Taking $p_2/k_2 \approx 5$,  we find that at merger the quadratic Love numbers contribute at the $\approx 10\%$  level to the tidal response relative to the linear Love numbers, with some dependence on the EoS and on the compactness.

Furthermore, as can be seen from Figs.~\ref{fig:phase} and~\ref{fig:cycles}, when the compactness is relatively small ($C\approx 0.1$) the quadratic tidal contribution is comparable to the next-to-next-to-leading-order linear tidal term entering at 7PN~\cite{Henry:2020ski}. This behavior also occurs for $C=0.15$ at high frequencies, whereas for $C=0.2$ the quadratic-Love contribution remains smaller than the 7PN linear tidal term. These findings are consistent with our expectations: owing to its stronger dependence on the compactness, the 8PN contribution from the quadratic Love number can become as important as the 7PN linear tidal contribution, depending on the stellar compactness.

Finally, from Fig.~\ref{fig:cycles}, we also see that, when the binary approaches the merger, the quadratic tidal contribution is typically much larger than the 4PN point-particle contribution, especially for small compactness.

%
\section{Conclusions}
\label{sec:conclusions}

In this work we have investigated the nonlinear tidal response of relativistic neutron stars by computing their quadratic Love numbers for realistic EoS and how these enter the GW phase from a binary coalescence.  
Our analysis combines the worldline EFT for extended gravitating objects with a direct solution of the Einstein equations up to second order in perturbation theory, allowing us to define and extract nonlinear tidal coefficients in a manner that is fully gauge-invariant.

We derived the second-order perturbation equations for realistic neutron-star models and computed the corresponding nonlinear tidal response for a representative set of nuclear-physics-motivated EoS.
The quadratic Love numbers exhibit a stronger scaling with the stellar radius than their linear counterparts, and are significantly enhanced in the small-compactness limit.  
This enhancement makes the quadratic tidal response potentially relevant for GW modeling. 

By matching the full relativistic solution to the worldline effective description, we extracted the quadratic tidal coefficient entering the binary dynamics and incorporated it into the conservative energy, GW flux, and resulting GW phase.

We found that, despite its nominally higher (8PN) order in the PN expansion, for typical neutron-star binaries the leading quadratic tidal term can be comparable to the next-to-next-to-leading-order linear tidal correction at 7PN, and can significantly exceed the 4PN point-particle contribution.
This behavior is especially pronounced for less compact configurations, where the enhancement of the quadratic response more than compensates for its higher PN suppression.

Dynamical Love numbers  also appear at 8PN order \cite{Chakrabarti:2013lua,Mandal:2023hqa,Combaluzier--Szteinsznaider:2025eoc, Kobayashi:2025vgl}. In contrast to nonlinear Love numbers, they exhibit logarithmic running due to gravitational nonlinearities, which is however less enhanced in the small-compactness limit.

Our results demonstrate that nonlinear tidal effects represent an essential ingredient for next-generation high-precision waveform modeling. State-of-the-art waveform models of neutron-star binaries, which are effectively informed also by high-order PN tidal corrections~\cite{Gamba:2020wgg,Abac:2023ujg,Haberland:2025luz,Abac:2025brd}, should include the quadratic tidal effects for better accuracy.
While such corrections are currently small, they will become relevant for future loud events, as those expected in third-generation ground-based detectors~\cite{ET:2025xjr}.
Interestingly, we also showed that a suitably normalized quadratic Love number can be written in term of the linear one using an approximate fitting formula that is mildly sensitive to the EoS. It would be interesting to explore this result further and check it against a larger set of EoS. Very recently, approximate universal relations have been discussed for the dynamical tidal coefficients of a neutron star~\cite{Saes:2025jvr}.

The formalism and numerical framework developed here pave the way for several further extensions, including the incorporation of magnetic nonlinear responses, dynamical tidal effects~\cite{Hinderer:2016eia,Steinhoff:2016rfi,Poisson:2020vap,Pitre:2023xsr,Katagiri:2024wbg,HegadeKR:2024agt,Andersson:2025iyd}, and higher-multipole nonlinear interactions.  
These directions will be crucial for constructing waveform models with the accuracy required to exploit the full scientific potential of upcoming GW observations.

\begin{acknowledgments}
We thank Sebastiano Bernuzzi, Valerio De Luca, Takuya Katagiri, and Andrea Valletta for discussions and checks.
PP is supported by the MUR FIS2 Advanced Grant ET-NOW (CUP:~B53C25001080001) and by the INFN TEONGRAV initiative. 
The research of LS has been funded, in part, by the French National Research Agency (ANR)
under project ANR-24-CE31-1097-01. 
MMR research is supported by the U.S. Department of
Energy (award no. DE-SC0011941).
This work has
received support under the program ``\textit{Investissement d'Avenir}'' launched by the French
Government and implemented by ANR, with the reference  ANR-18-IdEx-0001 as part of
its program ``\textit{Emergence}''.
  
\end{acknowledgments}
%


\appendix

\begin{widetext}

    \section{EFT one-point function}
\label{app:FDres}

Here we report the result for the $tt$-component of the EFT one-point function $\langle h_{\mu\nu}\rangle$, as obtained from the computation of the diagrams in Fig.~\ref{fig:FeynD}. We   write only the $\ell=2$ multipole, which is the one relevant for the matching. From the linear-response diagram in Fig.~\ref{fig:FeynD}(a), we obtain
\begin{align}
  \frac{\langle h_{tt}(r)\rangle_{\rm~\ref{fig:FeynD}(a)}^{\ell=2}}{2\Mpl}
  & = \mathscr{E}_{2,m}\frac{12 G \lambda_2}{r^3} \; .
\end{align}
Diagrams (b) and (c) together give the following contribution,
\begin{align}
  \frac{\langle h_{tt}(r)\rangle_{\rm~\ref{fig:FeynD}(b+c)}^{\ell=2}}{2\Mpl}
  & = \sum_{m_1 m_2} I^{m m_1 m_2}_{2 2 2} \mathscr{E}_{2,m_1} \mathscr{E}_{2,m_2} \frac{2 G \lambda_2}{r} \, .
\end{align}
Diagram (d) describes the nonlinear interaction between two linear responses and yields
\begin{align}
  \frac{\langle h_{tt}(r)\rangle_{\rm~\ref{fig:FeynD}(d)}^{\ell=2}}{2\Mpl}
  & = -\sum_{m_1 m_2} I^{m m_1 m_2}_{2 2 2} \mathscr{E}_{2,m_1} \mathscr{E}_{2,m_2}
      \frac{144 G^2 \lambda_2^{\,2}}{r^6} \, .
\end{align}
Finally, the nonlinear-response diagram (e) gives
\begin{align}
  \frac{\langle h_{tt}(r)\rangle_{\rm~\ref{fig:FeynD}(e)}^{\ell=2}}{2\Mpl}
  & = -\sum_{m_1 m_2} I^{m m_1 m_2}_{2 2 2} \mathscr{E}_{2,m_1} \mathscr{E}_{2,m_2}
      \frac{63 G \lambda_{222}}{2 r^4} \, .
\end{align}

\section{Angular integration}
\label{angIntApp}

For convenience, in this appendix we define $\Lambda_\ell \equiv \ell(\ell+1)$. To evaluate integrals of the product of two and three spherical harmonics---normalized such that $\int \D\Omega \, Y^{*}_{\ell m}(\theta,\varphi)Y_{\ell m}(\theta,\varphi) = 1$---and their derivatives over the sphere, we use the eigenvalue equation $\nabla_A\nabla^A Y_{\ell m} = -\Lambda_\ell Y_{\ell m}$, integration by parts, and the basic integral\footnote{Since the sphere has no boundary, we drop all total derivatives when performing integration by parts.}
\begin{equation}
    I^{m m_1 m_2}_{\ell \ell_1\ell_2} \equiv \int \D\Omega \,  Y^{*}_{\ell m}Y_{\ell_1 m_1}Y_{\ell_2 m_2}.
\end{equation}

Since the final result  depends on $m,m_{1}, m_2$ only through the symbol $I^{m m_1 m_2}_{\ell \ell_1\ell_2}$,  we will, for ease of notation, omit the $m,m_{1}, m_2$ labels when writing spherical harmonics in what follows.

From the Einstein equations we obtain at most two angular derivatives. After projecting out the linear equations we encounter the integrals
\begin{equation}
    \int \D\Omega \, \nabla_A Y_{\ell} \nabla^A Y_\ell =  \Lambda_\ell \,,
\end{equation}
and 
\begin{equation}
    \int \D \Omega\,  \nabla^{(A}\nabla^{B)} Y_{\ell} \nabla_{(A} \nabla_{B)} Y_\ell =  \frac{1}{2}\Lambda_\ell^2  - \Lambda_\ell \,,
\end{equation}
which are easily evaluated by moving all derivatives to one of the harmonics. At second order, we encounter
\begin{align}
    \int \D\Omega \, Y_\ell \nabla^A Y_{\ell_1} \nabla_A Y_{\ell_2} & =  \frac{1}{2}(\Lambda_{\ell_1} + \Lambda_{\ell_2} -\Lambda_\ell) I^{m m_{\ell_1} m_{\ell_2}}_{\ell \ell_1 \ell_2} \;, \\
    \int \D \Omega \, \nabla^{(A}\nabla^{B)} Y_{\ell} \nabla_{A} Y_{\ell_1} \nabla_{B} Y_{\ell_2} & =  \frac{1}{4} \left[ \Lambda _\ell \left(\Lambda _{\ell_1}+\Lambda _{\ell_2}\right)- \left(\Lambda _{\ell_1}-\Lambda _{\ell_2}\right){}^2\right] I^{m m_{\ell_1} m_{\ell_2}}_{\ell \ell_1 \ell_2} \;, \\
    \int \D\Omega \,  Y_{\ell_2} \nabla^{(A}\nabla^{B)} Y_{\ell_1} \nabla_{(A} \nabla_{B)} Y_\ell &= \frac{1}{4}  \left[ -2 \Lambda _{\ell_1} \left(\Lambda _{\ell_2}+1\right)+\Lambda _{\ell_1}^2+\left(\Lambda _{\ell_2}-\Lambda_\ell\right) \left(\Lambda _{\ell_2}-\Lambda
   _\ell+2\right)\right] I^{m m_{\ell_1} m_{\ell_2}}_{\ell \ell_1 \ell_2} \;.
\end{align}
The solution to the first integral can be derived starting from $\int \D \Omega \, Y_{\ell_1} Y_{\ell_2} \nabla^2 Y_\ell$, using the eigenvalue equation and integrations by parts. The rest of the integrals can be obtained  similarly.

\section{Particular solution and complete result for $p_{2}$}
\label{2ndOrderApp}

The full expression of the particular solution we used in the main text is
\begin{equation}
    H_{\rm part}(r) = I_{222}^{000}  \left[ F_0(r) + G \lambda_2 F_1(r) + G^2 \lambda_2^2 F_2(r) \right] \;, 
\end{equation}
where
\begin{align}
\label{partsol0}
F_{0}(r) &= -\frac{1}{4} r \left(2 r^3+r^2 r_s-3 r_s^3\right) \;, \\
    \label{partsol1}
    F_{1}(r) &= \frac{15 \left[ 672 r^5 r_s-1060 r^3 r_s^3+50 r^2 r_s^4+12 r^2 \left(r-r_s\right){}^2 \left(56 r^2+84 r r_s+19 r_s^2\right) \log
   \left(1-\frac{r_s}{r}\right)+117 r_s^6+160 r r_s^5\right]}{28 r \left(r-r_s\right) r_s^5} \;, \\
    F_{2}(r) & = -\frac{225}{r^2 \left(r-r_s\right)^2 r_s^{10}} \left[ 12 r r_s \left(48 r^5-105 r^4 r_s+58 r^3 r_s^2+13
   r^2 r_s^3-14 r r_s^4+2 r_s^5\right) \left(r-r_s\right) \log \left(1-\frac{r_s}{r}\right) \nonumber \right. \\ 
\label{partsol2}
  & \left.  +72 r^3 \left(4 r^2+6 r r_s-5 r_s^2\right) \left(r-r_s\right)^3 \log ^2\left(1-\frac{r_s}{r}\right) +r_s^2 \left(r_s-2
   r\right) \left(-144 r^5+630 r^4 r_s-780 r^3 r_s^2+195 r^2 r_s^3+95 r r_s^4+r_s^5 \right)\right] \;.
\end{align}


In addition, the complete expression for $p_{2}$  is 
\begin{equation}
    p_2 = \frac{8(1-2C)^{2} C^8}{245 D(C,y)^3} \left[z A(C) +  B(C)\log^2(1-2C) + L(C,y) \log(1-2C) + 2C\sum_{i = 0}^3 K_i(C) y^i\right] \;,
    \label{lambda222}
\end{equation}
where $A$ and $K_i$ are polynomials in $C$ and $B,D$ and $L$ are polynomials in $C$ and $y$. Their expressions are:
\begin{align}
    D(C,y) & = 2 C \left\{C \left[2 C (C (2 C (y+1)+3 y-2)-11 y+13)+3 (5 y-8) \right]-3 y+6 \right\}+3 (1-2 C)^2
   \log (1-2 C) [ 2 C (y-1)-y+2] \;, \\
    A(C) & = -7168 (1-2 C)^2 C^7 \;, \\
    B(C,y) & = 216 (1-2 C)^4 (2 C (y-1)-y+2)^3 \;, \\
    L(C,y) & =  9 (1-2 C)^2 (2 C (y-1)-y+2)^2 \nonumber \\ 
    & \times \{2 C [ 2 C (2 C (8 C (2 C (y+1)+3
   y-2)+157 y-141)-615 y+788)+687 y-1129]-245 (y-2) \} \;,
\end{align}
while the expressions for $K_i(C)$ read
\begin{align}
    K_0(C) & = -9600 C^{10}+35968 C^9-44288 C^8+35712 C^7+291312 C^6-1125936 C^5 \nonumber \\ & \quad +1769544
   C^4-1486464 C^3+698016 C^2-172944 C+17640 \;, \\
    K_1(C) & = -9600 C^{10}+10688 C^9+41344 C^8-11776 C^7-914832 C^6+2932200 C^5 \nonumber \\ & \quad -4069944
   C^4+3043764 C^3-1279980 C^2+285876 C-26460 \;, \\
    K_2(C) & = 9600 C^{10}-19200 C^9+1632 C^8-49632 C^7+916272 C^6-2491776 C^5 \nonumber \\
    & \quad +3046572
   C^4-2032164 C^3+769698 C^2-156168 C+13230 \;, \\
       K_3(C) & = 9600 C^{10}+2496 C^9-7008 C^8+28176 C^7-297552 C^6+689208 C^5 \nonumber \\
       & \quad -741996
   C^4+443154 C^3-152106 C^2+28233 C-2205 \;.
   \end{align}

\end{widetext}


\bibliographystyle{utphys}
\bibliography{biblio}

\end{document}